\documentclass[a4paper,11pt]{article}
\usepackage{graphicx} % Required for inserting images
\usepackage{subcaption}

\pdfoutput=1 % if you are submitting a pdf latex (i.e. if you have
\usepackage{jcappub} % for details on the use of the package, please
\usepackage{physics}
\usepackage{amsmath}     % For align environment
\usepackage{amssymb}     % For additional mathematical symbols
\usepackage{bm}
\usepackage{float}

\newcommand{\Dird}{\delta_{\rm D}^{[3]}}
\newcommand{\kron}{\delta^{\rm K}}

\title{Enabling Polyspectra via Stochastic Realizations of Inverse Decay Curvature Perturbations in Axion Inflation}
\author{Matthew A. Reinhard, Zachary Slepian, and Atsushi Taruya}
\date{September 2025}

\abstract{Current and upcoming galaxy-redshift surveys offer a promising path for constraining inflationary physics. These constraints are often imposed by relating correlations of the primordial curvature perturbation to the late-time galaxy correlation functions. The primordial correlators can either be computed analytically, or via realizations of the primordial curvature perturbations that are initialized on a grid and evolved numerically according to the classical equations of motion. In this work we take a hybrid approach, initializing the fields in the Bunch-Davies vacuum; and then imposing time evolution through direct numerical integration. Our method efficiently generates realizations of the inverse decay primordial curvature perturbations for axion inflation---an inflationary model in which a pseudoscalar inflaton field couples to a $U(1)$ gauge field via a Chern-Simons interaction. Depending on the choice of parameter $\xi\in\{2, 2.5, 3\}$, which controls gauge field particle production in this model, our method generates realizations with an average runtime of a few hours on 1 CPU core. We compute the power spectrum and bispectrum from these realizations and find a surprisingly large discrepancy between our result and previous work that used an approximation for the gauge field mode functions---up to a factor of $\sim4$ difference for the bispectrum in the case of $\xi=2$. This new method enables not only rapid computation of higher order polyspectra, but also generation of initial conditions for large-scale structure analyses.}         

\begin{document}

\maketitle

\section{Introduction}
The $\Lambda \rm{CDM}$ model has enjoyed tremendous phenomenological success in describing how small-scale inhomogeneities in the early Universe evolve over time, producing anisotropies in the Cosmic Microwave Background (CMB) and predicting the formation of large-scale structure (LSS) \cite{Frieman_2008}. The small-scale inhomogeneities in the primordial density field can be generated via quantum fluctuations of a scalar field in the very early Universe. If the potential energy of this scalar field dominates over its kinetic energy, rapid expansion of space ensues. This process is known as inflation, and the field responsible is known as the inflaton \cite{STAROBINSKY, Guth}.
In addition to inflation's solving the horizon and flatness problems \cite{Baumann}, inflationary fluctuations produce the nearly scale-invariant primordial power spectrum strongly favored by the CMB\cite{Bond_87, Planck_2018}, and set initial conditions for the growth of LSS \cite{Mukhanov}. \textit{The clustering of matter in LSS is thus directly related to the physics of inflation}.

The galaxy N-Point Correlation Functions (NPCFs) are central observables for extracting information from galaxy clustering data that may then be used to constrain inflationary models. In particular, the galaxy 4PCF is the simplest statistic sensitive to parity violation (breaking of mirror symmetry) \cite{Cahn_23}. Since General Relativity, a parity-symmetric theory, is primarily responsible for the formation and evolution of galaxies, any parity violation in LSS is likely due to new physics during inflation. The first measurements of the parity-odd galaxy 4PCF found tentative evidence for a nonzero signal \cite{Hou_2023, Philcox_2022}, though the large number of degrees of freedom (d.o.f) render the analysis very sensitive to uncertainties in the estimated covariance matrix. More recent measurements from Dark Energy Spectroscopic Instrument (DESI) \cite{slepian2025, Hou_2026} detected parity-odd modes at even higher statistical significance when performing the analogous analysis to \cite{Hou_2023}. However, they find that when performing a different analysis, in which the 4PCF measured from different, statistically independent patches of sky are cross-correlated, the putative signal is no longer visible \cite{Krolewski_2024}.  

Although the reliability of this detection of parity violation (PV) in LSS remains uncertain, we take it, or a potential future detection of PV,  as strong motivation for this work. Here we focus on a particular model of inflation known as axion inflation, in which a pseudoscalar inflaton field couples to a $U(1)$ gauge field via a Chern-Simons interaction \cite{Freese, Anber_2010, Barnaby_2011, Barnaby_2011_2}. This model dynamically breaks parity symmetry, producing a parity-odd contribution to the primordial trispectrum at the one-loop level. \cite{Niu_2023} demonstrated how to calculate this one-loop contribution and numerically computed the high-dimensional loop integral for two specific configurations of external wave vectors. \cite{reinhard2024} utilized the plane wave expansion and the Convolution Theorem to perform the angular integrals analytically, which significantly reduced the dimensionality of the loop integrals at the expense of introducing many infinite sums over the angular momenta. The search for this parity-violating signal in LSS will require computing the primordial trispectrum for a large number of wave vector configurations. 

 In axion inflation, the parity-violating ``inverse decay" curvature perturbations are due to the Chern-Simons interaction between the axion and gauge fields. In this work, we demonstrate a novel method for generating stochastic realizations of these inverse decay curvature perturbations. Each realization is computed on a grid in Fourier space by generating Gaussian realizations of the gauge field in Bunch-Davies vacuum, applying time evolution through deterministic mode functions, and evaluating the required integrals using numerical quadrature and Fast Fourier Transforms (FFTs). These realizations may then be used to efficiently calculate the NPCFs, and upon averaging this result over many realizations, the answer should approach the analagous analytical computation. 

The method presented here is similar in nature to previous lattice simulation studies \cite{Caravano_2023, Figueroa_2025, Jamieson_2025}, but is distinct in several respects. Whereas lattice simulations typically evolve the entire  inflaton field numerically according to the classical equations of motion, our code uses a perturbative approach, producing realizations of only the inverse decay inflaton perturbations at linear order. This allows us to work out the inverse decay power spectrum and bispectrum, and directly compare with previous analytical results \cite{Barnaby_2011, Durrer_2024}. Furthermore, we expect our approach to be significantly faster than numerically solving coupled second-order differential equations. It should be noted that the aformentioned lattice simulations are more general than the method presented here, taking into account backreaction i.e. the effect of the gauge field fluctuations on the inflaton background. Our approach neglects this effect. 

In Section \ref{Section2}, we provide a brief review of axion inflation, discussing the relevant physical processes and highlight the important equations. Whereas the vacuum perturbations are perfectly Gaussian, the inverse decay perturbations, sourced by a convolution of gauge field modes, are not. In Section \ref{Section3}, we outline our method for generating realizations of the inverse decay perturbations and discuss how to efficiently compute the momentum integral through FFTs and the time integral with Gaussian quadrature. Section \ref{Section4} details convergence tests; we ensure our sampling rate for the time integral is sufficient such that our discrete numerical implementation converges to the continuum result, and we confirm that our integral converges as the late-time bound approaches zero. In addition, we test convergence of our results in the UV and IR by measuring the power spectrum with varying Fourier grid sizes and box lengths. In Section \ref{Validation}, we validate the accuracy of our method by checking our results for the power spectrum and bispectrum against those of \cite{Barnaby_2011} and \cite{Durrer_2024}. In Section \ref{Section6}, we conclude.

As discussed in Section \ref{Section3}, we set the Fourier coefficients to zero for Nyquist modes. The vanishing of these modes directly follows from the fact that they must be transverse and also satisfy the polarization vector identities (\ref{pol_vec}); this statement is proven in Appendix \ref{AppendixA}. The method presented in this paper relies on the assumption that the gauge field is approximately classical. Appendix \ref{AppendixB} provides justification for these assumptions.

\section{Axion Inflation: A Brief Review}\label{Section2}
We consider here an inflationary model with a pseudoscalar inflaton field $\phi$ coupled to a $U(1)$ gauge field $A^\mu$ described by the following action:
\begin{align}
\mathcal{S} = \int \dd[4] x \sqrt{-g} \Bigg [ \frac{1}{2}M^2_{\rm p}R -\frac{1}{2} \partial_\mu \phi \, \partial^\mu \phi -\frac{1}{4}F_{\mu \nu} \,F^{\mu \nu} -\frac{1}{4\Lambda} \phi\, \tilde F^{\mu \nu}\, F_{\mu \nu} - V(\phi) \Bigg ],
\label{Action}
\end{align}
where $g$ is the determinant of the metric, $M_{\rm p}$ is the reduced Planck mass, $F_{\mu \nu} \equiv \partial_\mu A_\nu - \partial_\nu A_\mu $ is the gauge field strength tensor, $\tilde F_{\mu \nu} \equiv \epsilon^{\mu \nu \rho \sigma} F_{\rho \sigma}/2\sqrt{-g}$ is it's dual, $\Lambda^{-1}$ is the coupling constant, and $V(\phi)$ is the inflationary potential, which is assumed to be sufficiently flat to allow for slow roll inflation. 

During inflation, we assume a homogeneous and isotropic spacetime geometry given by the Friedmann--Lema\^{i}tre--Robertson–Walker (FLRW) metric:
\begin{align}
ds^2 = -dt^2 + a^2(t)dr^2 = a^2(\tau)[- d\tau^2 + dr^2], 
\end{align}
where we have set $c=1$. $a(\tau)$ is the scale factor, and we have chosen to work in conformal time $\tau$, which relates to physical time $t$ via $dt = a(\tau)d\tau$. In what follows, we neglect metric perturbations in the dynamical equations as these are sub-leading in slow-roll \cite{Barnaby_2011}. 

We now expand the inflaton field into a homogeneous background plus small perturbations:
\begin{align}
    \phi(\tau, \vb x) = \Bar \phi(\tau) + \delta \phi(\tau, \vb x).
\end{align}
We work in Coulomb gauge $\div{\vb A} = 0$ with $A_0=0$. The equations of motion for the inflaton and gauge field are then
\begin{align}
    &\phi '' + 2\mathcal{H}(\tau)\phi ' - \laplacian \phi + a(\tau)^2 \dv {V}{\phi} = \frac{a(\tau)^2}{\Lambda} \,\vb E \vdot \vb B \label{eom:phi} \hspace{10mm} \text{and}\\
    &\vb A'' - \laplacian \vb A - \frac{\bar{\phi}'(\tau)}{\Lambda}\grad \cross \vb A = 0, \label{eom:A}
\end{align}
where the electric and magnetic fields $\vb E$ and $\vb B$ are
\begin{align}\label{EM_Fields}
    &\vb E = -{a^{-2}(\tau)} \vb A' \hspace{10mm} \text{and} \nonumber\\
    &\vb B = {a^{-2}(\tau)}\, \curl \vb A.
\end{align}
The prime denotes a derivative with respect to conformal time, and $\mathcal{H(\tau)} \equiv a'(\tau)/a(\tau)$ is the conformal Hubble parameter, which is related to the standard Hubble parameter $H$ by $\mathcal{H} = aH$. 
The position-space gauge field can be written as an inverse Fourier transform: 
\begin{align}
    \label{A:ansatz}
    \vb A(\tau, \vb x) = \sum_{\lambda = \pm } \int \frac{\dd[3] \vb k}{(2\pi)^{3}} \left [ \bm{\hat{\epsilon}}_\lambda (\vu k) a_\lambda (\vb k) A_\lambda (\tau, k) e^{-i \vb k \vdot \vb x} + \text{h.c.} \right ],
\end{align}
where the circular polarization vectors $\bm{\hat{\epsilon}}_\lambda (\vu k)$ satisfy 
\begin{align}
\vb k \cdot \bm{\hat{\epsilon}}_{\pm} (\vu k) = 0, \hspace{2mm} \vb k \times \bm{\hat \epsilon}_{\pm}(\vu k) = \mp ik\bm{\hat \epsilon}_{\pm}(\vu k), \hspace{2mm} \bm{\hat\epsilon}_{\pm}(-\vu k) = \bm{\hat\epsilon}_{\pm}^*(\vu k),
\label{pol_vec}
\end{align}
h.c. stands for ``Hermitian conjugate", and the $^*$ represents a complex conjugate. $a_\lambda (\vb k)$ is the annihilation operator, which satisfies the canonical commutation relations 
\begin{align}
&\comm{a_\lambda(\vb k)}{a_{\lambda'}^\dag(\vb k')} = (2\pi)^3 \delta^{\rm K} _{\lambda \lambda'}\Dird(\vb k - \vb k')\hspace{5mm} \text{and} \\
&\comm{a_\lambda(\vb k)}{a_{\lambda'}(\vb k')} = \comm{a_\lambda^{\dagger}(\vb k)}{a_{\lambda'}^\dag(\vb k')}=0,
\end{align}
where $a_{\lambda}^\dag(\vb k)$ is the Hermitian conjugate of $a_{\lambda}(\vb k)$, and $\kron_{\lambda \lambda'}$ is the Kronecker delta. Inserting (\ref{A:ansatz}) into (\ref{eom:A}), we have the dynamical equation for the gauge field mode functions:
\begin{align}
\label{eom:Apm}
    A_{\pm}^{''}(\tau, k) + k^2A_{\pm}(\tau, k) \mp \frac{2k\xi}{\tau}A_{\pm}(\tau, k) = 0,   
\end{align}
where $$\xi \equiv \frac{\dot{\Bar{\phi}}}{2\Lambda H}$$
controls gauge field particle production. Throughout this work we assume $\xi$ is approximately constant and neglect its time dependence.

The $+$ mode is suppressed, whereas the $-$ mode experiences a tachyonic instability \cite{LargeNG}. From now on we only consider the tachyonically amplified mode, the solution for which is
\begin{align} \label{A_m}
    A_{-}(\tau, k) = \frac{1}{\sqrt{2k}}\bigg[G_0(\xi, -k\tau) + iF_0(\xi, -k\tau)\bigg],
\end{align}
where $F_0$ and $G_0$ are the Coulomb functions of the first and second kind, respectively.\footnote{https://mathworld.wolfram.com/CoulombWaveFunction.html}
An approximation for these mode functions, valid in the region $1/(8\xi) \lesssim -k\tau \lesssim 2\xi$, is 
\begin{align} \label{A_approx}
    A_{-}(\tau, k) \approx \frac{1}{\sqrt{2k}}\left(\frac{-k\tau}{2\xi}\right)^{1/4} e^{\pi \xi - 2\sqrt{-2\xi k\tau}}.
\end{align}
This approximation is extensively used to render calculations of the primordial correlators analytically tractable \cite{Anber_2010, Barnaby_2011_2, Barnaby_2011,  Sorbo_2011, Barnaby_2012, Cook_2012, Anber_2012, Linde_2013, Meerburg_2013, Ferreira_2014, Garc_a_Bellido_2016, Domcke_2017, _zsoy_2023, alam2024, Corba_2026}. In Section \ref{Validation}, we find a significant discrepancy between using the Coulomb mode functions and the approximation (\ref{A_approx}) when computing the power spectrum and bispectrum. 

We now highlight two important effects due to the interaction between the axion and the gauge field. First, the motion of the inflaton field background produces copious amounts of gauge quanta. Since one polarization is amplified and the other is suppressed, parity symmetry is broken. Second, these gauge quanta may recombine to produce perturbations in the inflaton field; this is commonly called inverse decay. These inverse decay perturbations will introduce corrections to the power spectrum and higher-order polyspectra. A third effect may also arise in which the gauge field may backreact and alter the homogeneous dynamics of the inflaton field. The importance of backreaction depends on the strength of the axion-gauge field coupling. In this work, we assume that backreaction is negligible.

The inflaton perturbation can be separated into those due to vacuum fluctuations and those due to inverse decay:
\begin{equation}
\delta \phi (\tau, \vb x) = \delta \phi^\text{vac} (\tau, \vb x) + \delta \phi^\text{inv} (\tau, \vb x).    
\end{equation}
These two types of perturbations are statistically independent \cite{Barnaby_2011}. In this work we focus only on the inverse decay perturbations; they obey the equation of motion
\begin{equation}
 \Bigg[\dv[2]{}{\tau} + 2\mathcal{H}(\tau)\dv[]{}{\tau} - \laplacian + a(\tau)^2 \dv [2] {V}{\bar\phi}\Bigg] \delta \phi^{\rm inv }  = \frac{a(\tau)^2}{\Lambda} \Big ( \vb E \vdot \vb B - \expval{\vb E \vdot \vb B} \Big ).
\label{eq:fluc_eom}
\end{equation}
As was done with the gauge field, we write the position-space inflaton perturbations as an inverse Fourier transform:
\begin{align} \label{Q_MF}
\delta \phi ^{\rm inv}(\tau, \vb x) = \int \frac{\dd [3] \vb k}{(2\pi)^{3}}\,\frac{Q^{\rm inv}(\tau, \vb k)}{a(\tau)} e^{-i\vb k \vdot \vb x}.
\end{align}
The inverse decay mode functions $Q^{\rm inv}$ are sourced by the gauge field:
\begin{align}\label{inv_mf}
    Q^{\text{inv}}(\tau, \vb k) = \int_{-\infty}^{0} \dd \tau' \,G(\tau, \tau', k)\,J(\tau', \vb k),
\end{align}
with the source 
\begin{equation} \label{J_def}
J(\tau, \vb k) \equiv \frac{a(\tau)^3}{\Lambda} \int \dd [3] \vb x\; \vb E \vdot \vb B \, e^{i \vb k \vdot \vb x}.  
\end{equation}
The effect of the last term on the right-hand side of Eqn. (\ref{eq:fluc_eom}) is to remove the zero mode contribution to the source. We explicitly remove this mode when constructing the source in Section \ref{Section3}. 

We are interested in modes which have been stretched outside the horizon. We can therefore take the late-time limit and obtain a simple expression for the Greens function\footnote{We believe there is a sign error in \cite{reinhard2024}, and have included an extra minus sign in the Green's function.}\cite{reinhard2024}:
\begin{align} \label{Green_function_limit}
G(\tau, \tau', k) = -\Theta(\tau - \tau') \frac{a(\tau) H \tau'}{k} (-k\tau)^{\frac{n_s - 1}{2}} j_1(-k\tau'),    
\end{align}
where $n_s$ is the scalar spectral index.

\section{The Method}\label{Section3}
In this section we demonstrate how to generate realizations of the inverse decay curvature perturbation in axion inflation through direct numerical integration. The curvature perturbation on uniform-density hyper-surfaces is given by
\begin{align}
\zeta^{\rm inv}(\tau, \vb k) = -\frac{H}{a(\tau) \dot{\bar{\phi}}} \; Q^{\rm inv}(\tau, \vb k).    
\end{align}
After canonical quantization, the field perturbations are quantum operators. For our method, we replace these operators with classical realizations of the fields. These classical realizations can then be used to compute correlation functions, and upon averaging over a large number of realizations, our result should match the analytical approach. We provide further details and a justification of this approach in Appendix \ref{AppendixB}. 

\subsection{Inverse Decay Realizations}
We begin by treating the gauge field as a classical random variable, from which we construct the classical source $J$. Assuming Bunch-Davies initial conditions, each mode obeys Gaussian statistics at early times. The time evolution of these fields is then encoded in the mode function. The tachyonic instability leads to huge production of gauge field particles; this amplification occurs for $|k\tau| \lesssim 2\xi$. We therefore take $|k\tau| = 2\xi$ as the UV cutoff for our time integration in Eq. (\ref{inv_mf}) \cite{Barnaby_2011, Niu_2023, Niu_2023_2}. The power spectrum has been found to be quite sensitive to the UV cutoff for $\xi \lesssim 2$ and not at all sensitive for $\xi > 2$ \cite{Durrer_2024}. In Figure \ref{real_vs_imaginary} of Appendix \ref{AppendixB}, we show that imaginary part of the gauge field mode functions is negligible when compared to the real part for most of the region $|k\tau| < 2\xi$. We therefore include only the real part of the mode functions in our calculations. In this approximation,
\begin{align}
    \label{A_rm}
    \vb A(\tau, \vb x) &= \int \frac{\dd[3] \vb q}{(2\pi)^{3}} \bm{\hat{\epsilon}}_- (\vu q) [a_- (\vb q) + a^\dag_- (\vb{-q})] A_- (\tau, q) e^{-i \vb q \vdot \vb x} \nonumber\\
    & \to \int \frac{\dd[3] \vb q}{(2\pi)^{3}} \vb{X_-}(\vb q) A_- (\tau, q) e^{-i \vb q \vdot \vb x},
\end{align}
where we have used the last property of the polarization vector listed in (\ref{pol_vec}), and we have included only the tachyonic $A_-$ modes in the integral as discussed in Section \ref{Section2}. We have replaced the operators and the polarization vector with $\vb{X_-}(\vb q)$, which is a random 3D vector that satisfies the identities (\ref{pol_vec}). 

The source term can now be written as
\begin{align}
    J(\tau, \vb k) = -\frac{1}{\Lambda a(\tau)} \int \frac{\dd[3] \vb q}{(2\pi)^3} \; q \; X_-^i(\vb k - \vb q) X_-^i(\vb q) A'_- (\tau, |\vb k - \vb q|)  A_- (\tau, q),
\end{align}
where the superscript on $X_-^i(\vb q)$ denotes the $i$-th spatial component of the vector $\vb{X_-}(\vb q)$ and we have used (\ref{J_def}) along with the definitions of the electric and magnetic fields (\ref{EM_Fields}) and the second identity in (\ref{pol_vec}). The prime on the mode function denotes a derivative with respect to conformal time. Now, for the inverse decay curvature perturbation we have
\begin{align}\label{curvature}
    \zeta^{\rm inv}(\tau, \vb k) &= \frac{H}{\Lambda a(\tau) \dot{\bar\phi}} \int_{-\infty}^0 \frac{d\tau'}{a(\tau')} \; G(\tau, \tau', k) \int \frac{\dd[3] \vb q}{(2\pi)^3} \; q \; X_-^i(\vb k - \vb q) X_-^i(\vb q) A'_- (\tau', |\vb k - \vb q|)  A_- (\tau', q) \nonumber \\
    &\equiv \frac{H}{\Lambda a(\tau) \dot{\bar \phi}} \int_{-\infty}^0 \frac{d\tau'}{a(\tau')} G(\tau, \tau', k) S(\tau', \vb k),    
\end{align}
where we have defined $S(\tau', \vb k) = -\Lambda \, a(\tau')\, J(\tau', \vb k)$ for later convenience.
During inflation $a(\tau') \approx -1/(H \tau')$, thus we have
\begin{align}\label{Final_curvature}
    \zeta^{\rm inv}(\tau, \vb k) &= 8\pi^2 \xi \,  (-k\tau)^{\frac{n_s - 1}{2}}\, \mathcal{P} \int_{-\infty}^0 d\tau' \, \mathcal{K}(\tau, \tau', k) \, S(\tau', \vb k) \nonumber\\
    &= 8\pi^2\, \xi \, (-k\tau)^{\frac{n_s - 1}{2}}  \mathcal{P} \, R(\tau, \vb k),
\end{align}
where we have defined the kernel $\mathcal{K}$ as 
\begin{align} \label{Green_function}
\mathcal{K}(\tau, \tau', k) \equiv  \frac{\tau'^2}{k} \, j_1(-k\tau'),    
\end{align}
so that our curvature perturbation can be written as the stochastic response 
\begin{align}\label{Response}
 R(\tau, \vb k) \equiv \int_{-\infty}^0 d\tau' \, \mathcal{K}(\tau, \tau', k) \, S(\tau', \vb k),  
\end{align}
multiplied by the dimensionless quantities $\xi$,
\begin{align}
    \mathcal{P} \equiv \bigg(\frac{H^2}{2\pi|\dot{\bar\phi}|}\bigg)^2 ,
\end{align}
and $(-k\tau)^{(n_s - 1)/2}$.

We note that $S(\tau', \vb k)$ is a convolution, so we may use the Convolution Theorem to obtain
\begin{align}\label{Source}
    S(\tau', \vb k) = \mathcal{F}\Bigg[\mathcal{F}^{-1}\bigg[X_-^i(\vb q) A'_- (\tau', q)\bigg](\vb x) \; \mathcal{F}^{-1}\bigg[qX_-^i(\vb q) A_- (\tau', q)\bigg](\vb x) \, \Bigg](\vb k),
\end{align}
where $\mathcal{F}$ and $\mathcal{F}^{-1}$ denote the Fourier transform and its inverse, respectively. We utilize Fast Fourier Transforms (FFTs) at each point in time $\tau'$, enabling computation of $S(\tau', \vb k)$ in $\mathcal{O}(N_g \log N_g)$, where $N_g$ is the size of the Fourier grid. Thus the curvature perturbation computation is $O(N_t \, N_g \log N_g)$, where $N_t$ is the number of sample points for the time integration.

\subsubsection{Discretization Scheme} \label{Discretization}
We initialize a cubic grid in configuration space with $N_g$ points. Assuming that the gauge field has a vanishing vacuum expectation value and begins in the Bunch-Davies vacuum, we initialize the gauge field modes by drawing a real 3D vector from a white noise generator at each grid point. Each vector component is randomly generated from a Gaussian distribution with zero mean and variance set by the grid spacing $\sigma^2 =(N/L)^{3}$ \footnote{This corresponds to unit variance in the continuum.}, where $L$ is the side length of the cubic grid and $N = N_g^{1/3}$. 

Next, we Fourier transform this vector field and project it onto the $-$ helicity direction with the operator 
\begin{align}
    P^-_{ij} = \frac{1}{2}(\delta_{ij} - \hat k_i \hat k_j + i\epsilon_{ijl} \hat k_l ).
\end{align}
This ensures that the vector field satisfies the conditions (\ref{pol_vec}). The fact that we generated a real vector field in configuration space also ensures that our Fourier space field satisfies the reality condition
\begin{align}\label{reality}
    \vb X_-^{*}(-\vb q) = \vb X_-(\vb q).
\end{align}
On a discrete grid, the Fourier modes are given by
\begin{align}
    \vb k = \frac{2\pi}{L}(n_x, n_y, n_z) \equiv \frac{2\pi}{L}\vb n .
\end{align}
The Nyquist wavenumber 
\begin{align}
    k_{Ny} = \frac{\pi N}{L}
\end{align}
is the largest wavenumber that can be resolved on the grid. For wave vectors $\vb{p}$ with at least one component equal to the Nyquist wavenumber, the only way that both (\ref{pol_vec}) and (\ref{reality}) can be satisfied is if $\vb X_-(\vb p) = 0$; we demonstrate this fact in Appendix \ref{AppendixA}. We thus set these modes to zero on our grid.

As dictated by (\ref{Source}), at each sampled moment in time $\tau'$, we multiply our stochastic field $\vb X_-(\vb q)$ by the deterministic mode function $A_- (\tau', q)$ or its derivative. We then enforce the UV cutoff by setting to zero any modes which violate the condition $|k\tau'| \leq 2\xi$.  We now express the Coulomb functions in terms of the confluent hypergeometric functions of the second kind $U(a,b,c)$ \cite{Gaspard_2018}
\begin{align}
    H_0^{(+)}(\xi,\rho) &= G_0(\xi, \rho) + iF_0(\xi, \rho), \nonumber\\
    & =-2ie^{\frac{\pi \xi}{2}}e^{i\sigma_0}\rho \,e^{i\rho} \; U(1 + i\xi, 2, -2i\rho)
\end{align}
where
$$\sigma_0 \equiv \arg \Gamma(1+i\xi),$$
and $\Gamma(x)$ is the Gamma function\footnote{https://mathworld.wolfram.com/GammaFunction.html}. The derivative of the Coulomb functions can then be calculated analytically using the recursion relation\footnote{https://dlmf.nist.gov/13.3}
\begin{align}
    \dv{}{z}U(a,b,z) = -a \;U(a+1,b+1,z).
\end{align}
To implement the Coulomb mode functions we use the \texttt{mpmath} library's \texttt{hyperu} function and take the real part; we neglect the subleading imaginary part as discussed in Appendix \ref{AppendixB}. We use cubic spline interpolation to approximate the mode function and its derivative. Repeatedly calling this interpolator at each sampled time is significantly faster than doing the same with \texttt{hyperu}.

\subsubsection{Anti-Aliasing}
The two inverse Fourier Transforms in (\ref{Source}) can be written as 
\begin{align}
    \int \frac{\dd [3] q_1}{(2\pi)^3} \int \frac{\dd [3] q_2}{(2\pi)^3} \; [\vb X_-(\vb q_1) \vdot \vb X_-(\vb q_2)]\; q_2 \; A'_- (\tau', q_1) A_- (\tau', q_2)\; e^{-i(\vb q_1 + \vb q_2) \vdot \vb x} 
\end{align}
On our discrete Fourier grid this becomes
\begin{align}
    \frac{1}{L^6}\sum_{\vb m} \sum_{\vb n} \; [\vb X_-(\vb q_1) \vdot \vb X_-(\vb q_2)]\; q_2 \; A'_- (\tau', q_1) A_- (\tau', q_2) \; e^{-\frac{2\pi i}{L}(\vb m + \vb n) \vdot \vb x}
\end{align}
Although individually the components of $\vb m$ and $\vb n$ are less than or equal to the Nyquist bound of $N/2$, their sum in general is not. These higher frequency modes cannot be resolved by our grid; their complex amplitudes are incorrectly attributed to lower frequency modes. This phenomenon is known as aliasing.

We implement a 3/2 anisotropic de-aliasing scheme in which we extend the size of our Fourier grid along each dimension to $3N/2$ and set the Fourier coefficient for each of these new modes to zero \cite{dealising, Taruya_2018, Taruya_2021, Taruya_2022}. This grid extension is performed before taking the two inverse Fourier transforms in (\ref{Source}). When the two terms are multiplied in real space, any aliasing then affects only the newly added Fourier modes. After performing the forward Fourier transform, we then return our grid to its original size, removing these aliased modes. We then remove the zero mode of the source by setting $S(\tau', \vb 0) = 0$, as discussed in Section \ref{Section2}.

\subsubsection{Time Integration} \label{time_int}

We perform the time integration for Eq. (\ref{curvature}) in log space using Legendre-Gauss quadrature. This is implemented using the \texttt{numpy} library's \texttt{leggauss} function. Legendre-Gauss quadrature is performed on an integration region of $[-1,1]$; we therefore must perform a change of interval. First, we express our integration in log space:
\begin{align} \label{int_f(x)}
 \int_{a}^{b} f(x)\dd x = \int_{\ln a}^{\ln b} f(e^{u}) e^u \dd u  
\end{align}
where $u \equiv \ln x$. Furthermore, we define $r \equiv \frac{1}{2} \ln(b/a)$ and $s \equiv \frac{1}{2}\ln(ab)$ so that $u = r\eta + s$. Changing the variable of integration to $\eta$, (\ref{int_f(x)}) now becomes
\begin{align}
\int_{-1}^{1} f(e^{r\eta + s}) e^{r\eta + s} \dv []{u}{\eta} \dd \eta = \int_{-1}^{1} f(x(\eta)) x(\eta) r \dd \eta 
\end{align}
On a discrete grid this integral becomes
\begin{align}
\sum_i^{N_{\rm t}}w_if(x(\eta_i)) x(\eta_i) r  
\end{align}
where the sample points $\eta_i$ and weights $w_i$ are determined by \texttt{leggauss}.

As previously discussed, we impose a UV cutoff for time integration such that $-k \tau' < 2\xi$. This conditions implies that the early-time bound depends on the wavenumber $k$. Rather than adaptively changing the integration bound for every point on the Fourier grid, we instead choose a global early-time integration bound of $\tau'_{\rm min} = -2\xi/k_{\rm min}$. This ensures that all modes in the tachyonic region are retained. Although this global bound will also include modes outside the tachyonic region, we recall from Section \ref{Discretization} that we have already set these modes to zero, so they will not contribute to the integral.

\section{Convergence Tests}\label{Section4}

For our convergence tests, we work with dimensionless quantities which are related to the physical quantities by an arbitrary reference scale $k_*$,
\begin{align}
    \boldsymbol{\kappa} = \frac{\vb k}{k_*}, \hspace{5mm}  \tilde L = k_* L, \hspace{5mm} \tilde \tau = k_* \tau.  
\end{align}
Here $\kappa$ is the dimensionless wavenumber, $\tilde L$ is the dimensionless side length of our cubic grid, and $\tilde \tau$ is the dimensionless conformal time. For a dimensionless box length of $\tilde L = 2\pi$, we have the nice relation
\begin{align}
    \boldsymbol{\kappa} = \frac{2\pi \vb n}{\tilde L} = \vb n. 
\end{align}
The UV and IR bounds for our Fourier transforms are set by 
\begin{align}
    &\kappa_{Ny} = \frac{N\pi}{\tilde{L}} \hspace{10mm} \text{and}  \hspace{10mm} \kappa_{\rm IR} = \frac{2\pi}{\tilde{L}},
\end{align}
respectively. We note that the IR bound also determines the grid resolution $\Delta \kappa$. 

We will use the power spectrum as a diagnostic to test convergence of the integrals in Eq. (\ref{curvature}). The power spectrum is evaluated at some late time $\tau$ towards the end of inflation, we will therefore neglect any time dependence of the functions for what follows. We define the inverse decay power spectrum $P^{\rm inv}_{\zeta}(k)$ by
\begin{align}
    \expval{\zeta^{\rm inv}(\vb k_1) \zeta^{\rm inv}(\vb k_2)} = (2\pi)^3P^{\rm inv}_{\zeta}(k) \, \Dird(\vb k_1 + \vb k_2),
\end{align}
where $k = |\vb k_1| = |\vb k_2|$, and the response power spectrum $P_R(k)$ by 
\begin{align}
    \expval{R(\vb k_1) R(\vb k_2)} =  (2\pi)^3P_{R}(k) \, \Dird(\vb k_1 + \vb k_2),
\end{align}
where the stochastic response $R(\tau, \vb k)$ was defined in (\ref{Response}). These two quantities are related:
\begin{align}\label{P_inv}
    P^{\rm inv}_{\zeta}(k) = \bigg [8\pi^2 \xi (-k\tau)^{\frac{n_s-1}{2}} \mathcal{P}\bigg]^2 P_{R}(k).
\end{align}
We will find it convenient to work with the dimensionless power spectra:
\begin{align} \label{Dimensionless_PS}
    \Delta_{\zeta}^{2 \; \rm inv}(k) \equiv \frac{k^3}{2\pi^2} P^{\rm inv}_{\zeta}(k) \hspace{5mm} \text{and} \hspace{5mm} \Delta^2_{R}(k) \equiv \frac{k^3}{2\pi^2} P_{R}(k).
\end{align}
We choose $\xi \in \{2,2.5,3\}$ for the following tests as it facilitates direct comparison with previous studies (see Section \ref{Validation}). All power spectra are computed with the \texttt{Pk} function from the \texttt{Pylians3} library; we retain only the monopole term.

\subsection{Time Integral Convergence} \label{Time Convergence}
The late-time bound for the time integral in (\ref{curvature}) is zero, but our numerical implementation replaces this bound with some small nonzero value. We check that the integral converges in the limit as the late-time bound approaches zero by generating realizations of the response $R(\vb k)$, with grid size $N_g=256^3$, and dimensionless box length $\tilde L = 2\pi$, each with a different late-time integration bound\footnote{When using Legendre-Gauss quadrature to compute the integral, the integrand is never actually evaluated at these boundary points. Nonetheless, decreasing the late-time bound in this way does effectively decrease the smallest sampled value.}. We then compare these results against a reference using the relative $\ell^2$-norm:
\begin{align}
    \text{error} = \frac{\norm{R(\vb k) - R_{\rm ref}(\vb k)}_2}{\norm{R_{\rm ref}(\vb k)}_2},
\end{align}
where the $\ell^2$-norm of an $N$-dimensional complex vector $\vb A$ is
\begin{align}
    \norm{\vb A}_2 \equiv \sqrt{\sum_{i}^N|A_i|^2}.
\end{align}
The reference is computed with a late-time bound of $\tilde \tau = -0.0075$. The results are shown in Table \ref{tab:late-time-convergence}. We find that a late-time bound of $\tilde{\tau} = -0.09$ has an acceptable relative error of $\sim0.4\%$. For the rest of this work we use this bound for all time integrals.

Additionally, we must ensure that the integral sampling rate is sufficient such that our discrete implementation converges to the continuum result. We compute the response $R$ with the same random seed and vary the number of time integration points $N_{\rm t}$ for $\xi \in \{2,2.5,3\}$. These realizations are all computed on an $N=256$ grid with $\tilde L = 2\pi$ and then compared against a reference via the $\ell^2$-norm. The results can be found in Table \ref{tab:time-convergence}. 

We find that the required number of time integration points rises steeply as $\xi$ decreases. We tested trapezoid rule against Legendre-Gauss quadrature for the case of $\xi=2$ and found that the former converges slightly faster. This appears to be due to the fact that the integrand (\ref{curvature}) becomes much less smooth as $\xi$ decreases. This occurs because for a given $\kappa$, our implementation of the UV cutoff only activates modes at time $|\tilde\tau| < 2\xi/\kappa$. For large values of $\xi$, these modes get activated when the integrand is still quite small compared to its maximum value. For smaller values of $\xi$, these modes get activated where the integrand has significant support, leading to sharp features in the integrand. For this reason, we use trapezoid rule for the case of $\xi=2$ and Legendre-Gauss quadrature for $\xi=2.5$ and $3$ throughout this work.

\begin{table}[htbp]
    \centering
    \small
    \renewcommand{\arraystretch}{1.15}
    \setlength{\tabcolsep}{5pt}

    \begin{minipage}[t]{0.26\textwidth}
        \centering
        $\xi=2$ \par\vspace{6pt}
        \begin{tabular}{|c|c|}
            \hline
            $|\tilde\tau_{f}|$ & Percent error \\
            \hline
            0.180 & $0.461\%$ \\
            0.120 & $0.466\%$ \\
            0.090 & $0.477\%$ \\
            0.060 & $0.452\%$ \\
            0.030 & $0.539\%$ \\
            0.015 & $0.502\%$ \\
            \hline
        \end{tabular}
    \end{minipage}\hspace{4pt}%
    \begin{minipage}[t]{0.27\textwidth}
        \centering
        $\xi=2.5$ \par\vspace{6pt}
        \begin{tabular}{|c|c|}
            \hline
            $|\tilde\tau_{f}|$ & Percent error \\
            \hline
            0.180 & $0.414\%$ \\
            0.120 & $0.435\%$ \\
            0.090 & $0.400\%$ \\
            0.060 & $0.428\%$ \\
            0.030 & $0.414\%$ \\
            0.015 & $0.444\%$ \\
            \hline
        \end{tabular}
    \end{minipage}\hspace{4pt}%
    \begin{minipage}[t]{0.25\textwidth}
        \centering
        $\xi=3$ \par\vspace{6pt}
        \begin{tabular}{|c|c|}
            \hline
            $|\tilde\tau_{f}|$ & Percent error \\
            \hline
            0.180 & $0.396\%$ \\
            0.120 & $0.306\%$ \\
            0.090 & $0.297\%$ \\
            0.060 & $0.309\%$ \\
            0.030 & $0.334\%$ \\
            0.015 & $0.356\%$ \\
            \hline
        \end{tabular}
    \end{minipage}

    \caption{Late-time convergence of the response $R$ for $\xi\in\{2,2.5,3\}$ is shown. For a single random seed, $R$ was computed with varying values of the late-time bound $\tilde \tau_{f}$ for the time integral in (\ref{curvature}). The reported error was obtained by comparing each $R$ against a reference using the relative $\ell^2$-norm as described in Section \ref{Time Convergence}. The reference values were computed with $|\tilde{\tau}| = 0.0075$, and each set of $R$ was computed with $N_t = 300, 100,$ and $50$, for $\xi = 2,2.5,$ and $3$, respectively. Trapezoid rule was use for the case of $\xi = 2$, and Legendre-Gauss quadrature for $\xi=2.5$ and $3$}
    \label{tab:late-time-convergence}
\end{table}

\begin{table}[H]
    \raggedright
    \small
    \setlength{\tabcolsep}{5pt}
    \hspace*{15mm}%
    \begin{minipage}[t]{0.25\textwidth}
        \centering
        $\xi = 2$\par
        \vspace{4pt}
        \begin{tabular}{|c|c|}
            \hline
            $N_{\mathrm{t}}$ & Percent error \\
            \hline
             70 & $1.224\%$ \\
             80 & $1.134\%$ \\
             90 & $0.966\%$ \\
            100 & $0.858\%$ \\
            110 & $0.793\%$ \\
            120 & $0.717\%$ \\
            130 & $0.657\%$ \\
            \hline
        \end{tabular}
    \end{minipage}\hspace{0.02\textwidth}%
    \begin{minipage}[t]{0.25\textwidth}
        \centering
        $\xi = 2.5$\par
        \vspace{4pt}
        \begin{tabular}{|c|c|}
            \hline
            $N_{\mathrm{t}}$ & Percent error \\
            \hline
            18 & $1.601\%$ \\
            20 & $1.423\%$ \\
            22 & $1.289\%$ \\
            24 & $1.149\%$ \\
            26 & $1.058\%$ \\
            28 & $0.993\%$ \\
            30 & $0.934\%$ \\
            \hline
        \end{tabular}
    \end{minipage}\hspace{0.02\textwidth}%
    \begin{minipage}[t]{0.25\textwidth}
        \centering
        $\xi = 3$\par
        \vspace{4pt}
        \begin{tabular}{|c|c|}
            \hline
            $N_{\mathrm{t}}$ & Percent error \\
            \hline
            12 & $1.242\%$ \\
            14 & $0.931\%$ \\
            16 & $0.749\%$ \\
            18 & $0.665\%$ \\
            20 & $0.592\%$ \\
            22 & $0.529\%$ \\
            24 & $0.504\%$ \\
            \hline
        \end{tabular}
    \end{minipage}

    \caption{Time-integration convergence for an $N=256$ grid
    with $\widetilde{L}=2\pi$ and a late-time bound of
    $\tilde{\tau}=-0.09$. We test convergence for $\xi=2,2.5,$
    and $3$ using reference values computed with
    $N_{\mathrm{t}}=300,100,$ and $50$, respectively.
    The trapezoidal rule was used for $\xi=2$, and
    Gauss--Legendre quadrature for $\xi=2.5$ and $3$.
    The error was computed using the relative $\ell^2$-norm
    as described in Section~\ref{Time Convergence}.}
    \label{tab:time-convergence}
\end{table}

\subsection{UV Convergence} \label{UV_section}

\begin{figure}[H]
    \centering
    \includegraphics[width=0.6\textwidth]{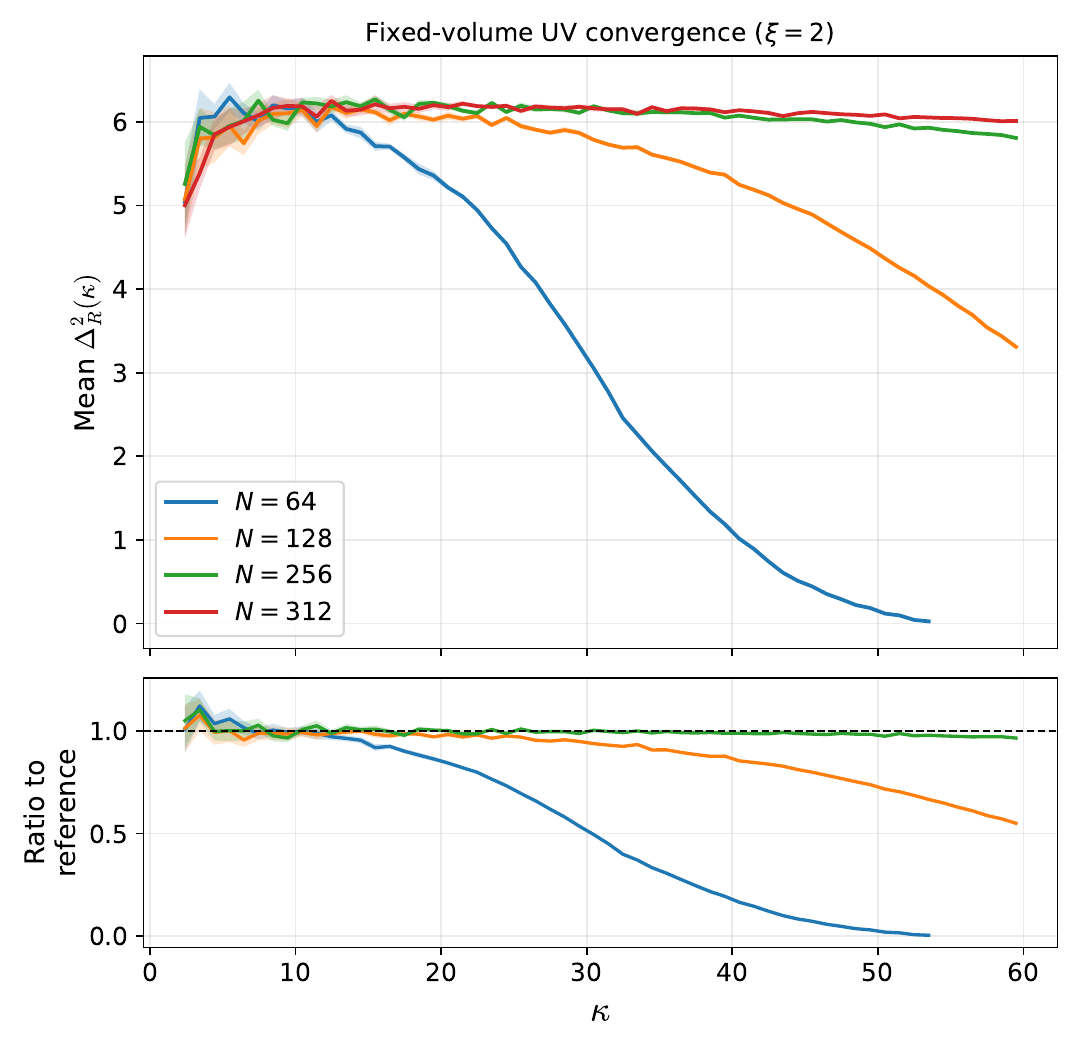}
    \caption{In the top panel, the dimensionless power spectrum of the inverse decay response $\Delta^2_{R}(\kappa)$ is shown for different grid sizes N $\in\{64, 128, 256, 312\}$ for the case of $\xi = 2$. For each grid size, 10 realizations were generated and their power spectra were computed using the \texttt{pylians3} library; we then average over the power spectra at each $\kappa$. The shaded region around each curve gives the $1 \sigma$ uncertainty of the mean. In the bottom panel, the ratio of the power spectra to the N = 312 reference is shown. The different curves agree around $\kappa = 10$, indicating that integral (\ref{curvature}) has converged for this value of $\kappa$. As $\kappa$ increases, the power spectrum of smaller N grids begins to dip due to their UV bound being insufficiently large for integral (\ref{curvature}) to converge. }
    \label{fig:UV-xi2}
\end{figure}

\begin{figure}[htbp]
    \centering
    \includegraphics[width=0.6\textwidth]{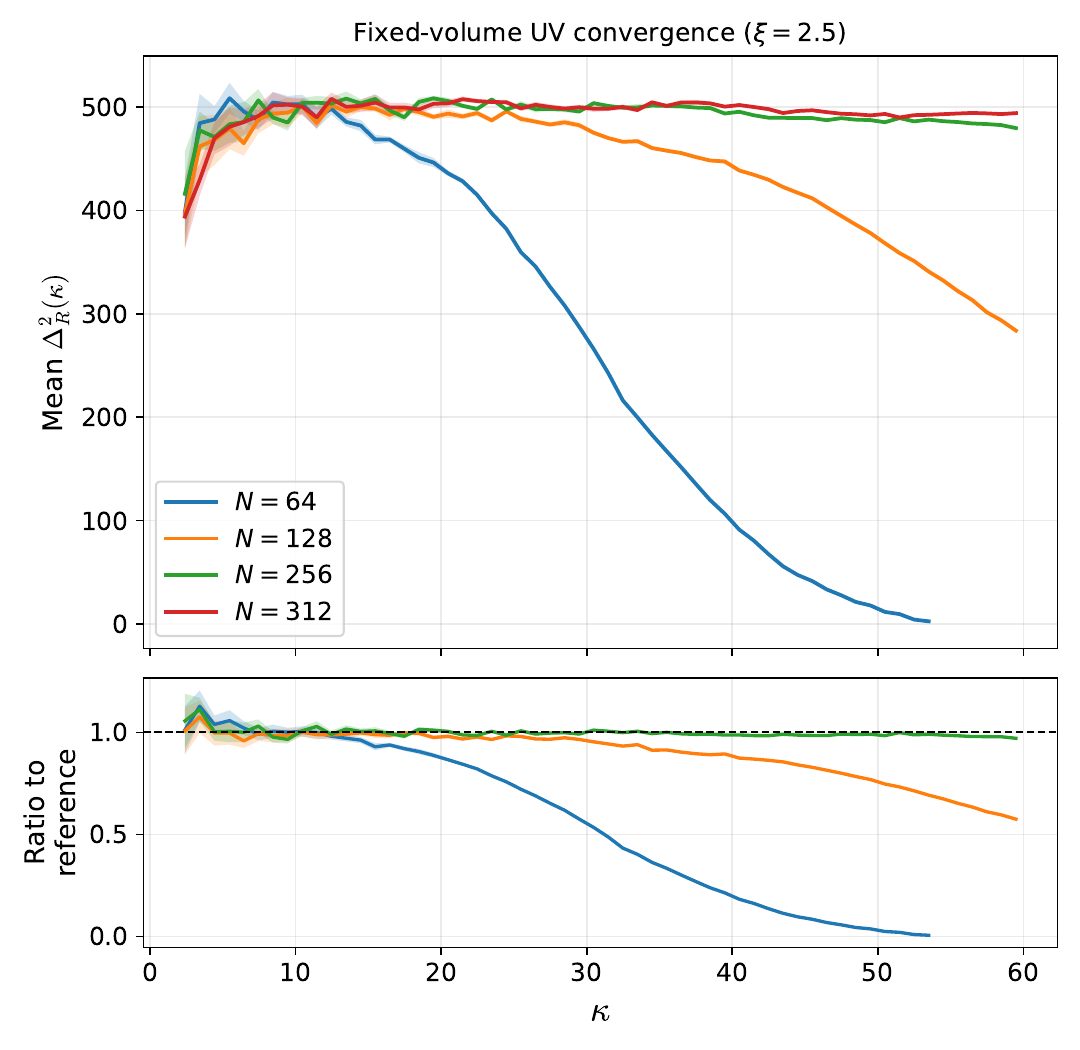}
    \caption{In the top panel, the dimensionless power spectrum of the inverse decay response $\Delta^2_{R}(\kappa)$ is shown for different grid sizes N $\in\{64, 128, 256, 312\}$ for the case of $\xi = 2.5$. See Figure \ref{fig:UV-xi2} and Section \ref{Section4} for more details.}
    \label{fig:UV-xi2p5}
\end{figure}

\begin{figure}[htbp]
    \centering
    \includegraphics[width=0.6\textwidth]{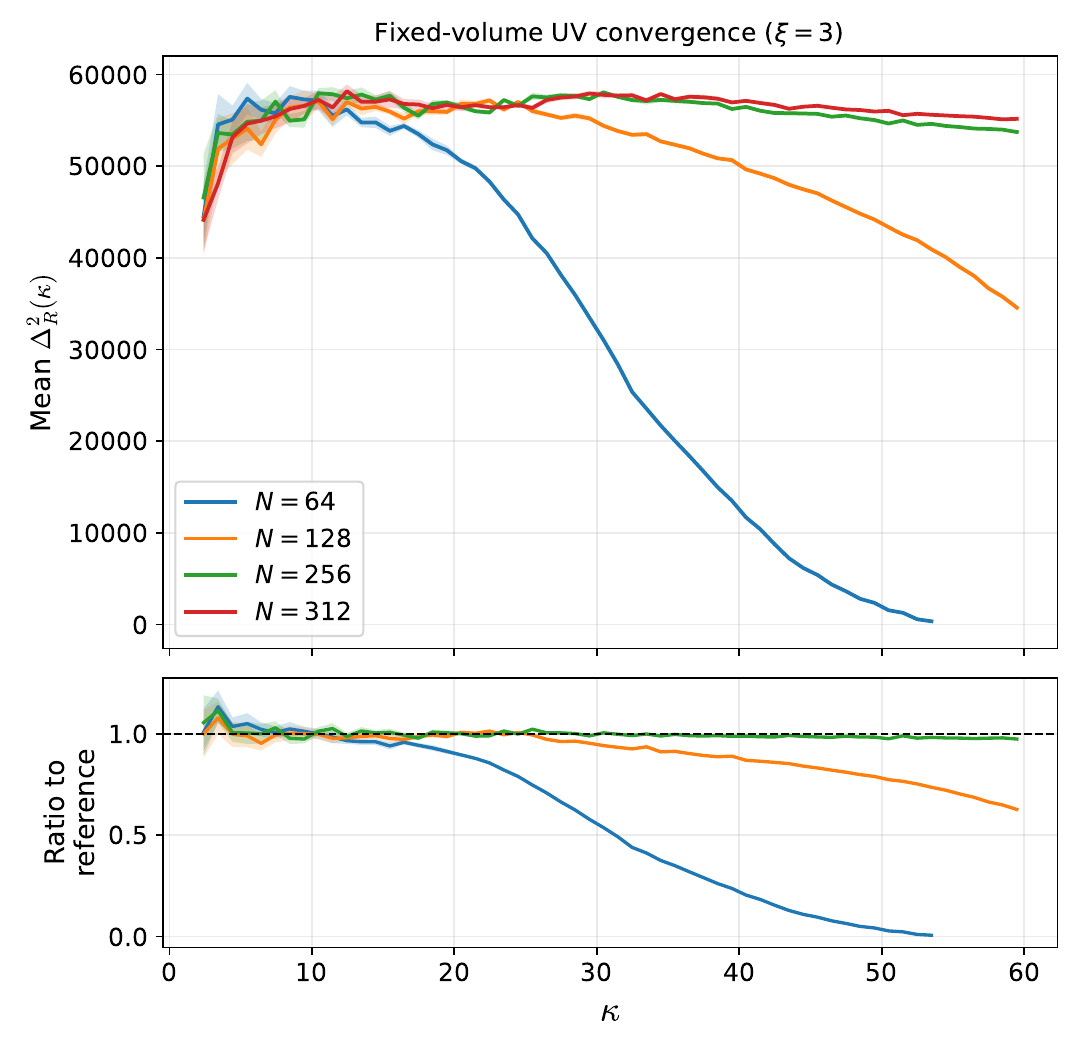}
    \caption{In the top panel, the dimensionless power spectrum of the inverse decay response $\Delta^2_{R}(\kappa)$ is shown for different grid sizes N $\in\{64, 128, 256, 312\}$ for the case of $\xi = 3$. See Figure \ref{fig:UV-xi2} and Section \ref{Section4} for more details.}
    \label{fig:UV-xi3}
\end{figure}

At fixed box length $\tilde L$, the size of our grid $N$ determines the largest $\vb q$ included in the momentum integral (\ref{curvature}) i.e. the UV bound. We must ensure that our grid size is sufficient for convergence of this integral. To this end, we hold the IR bound of the momentum integral fixed by setting $\tilde{L} = 2\pi$, and compute 10 realizations of the curvature perturbation for each grid size N $\in\{64, 128, 256, 312\}$, using the same random seeds for each set of 10 realizations. We compute the dimensionless power spectrum of the inverse decay response $\Delta^2_{R}(\kappa)$ for each realization and average them at each value of $\kappa$. The results are shown for $\xi \in \{2,2.5,3\}$ in Figures \ref{fig:UV-xi2}, \ref{fig:UV-xi2p5}, and \ref{fig:UV-xi3}.

$\Delta^2_{R}(\kappa)$ does not contain any scale-dependence; its spectrum should be flat over the converged $\kappa$ range, with some small deviations due to stochastic noise. In Figures \ref{fig:UV-xi2}, \ref{fig:UV-xi2p5}, and \ref{fig:UV-xi3}, we additionally see deviation from flatness for large and small $\kappa$; this deviation is due to two important numerical effects. For small $\kappa$, we approach the resolution of our discrete grid, meaning there are very few wave vectors with this magnitude. This leads to poor convergence of the momentum integral in (\ref{curvature}) and a significant underestimation of the power spectrum at these scales. As $\kappa$ increases, the number of modes required for integral (\ref{curvature}) to converge increases as well. A larger Nyquist frequency improves this convergence by effectively raising the UV cutoff for this integral. This is reflected in the figures; the plateau-like region of the power spectrum extends to larger $\kappa$ at larger grid sizes.    

\subsection{IR Convergence}\label{IR_section}

\begin{figure}[H]
    \centering
    \includegraphics[width=0.6\textwidth]{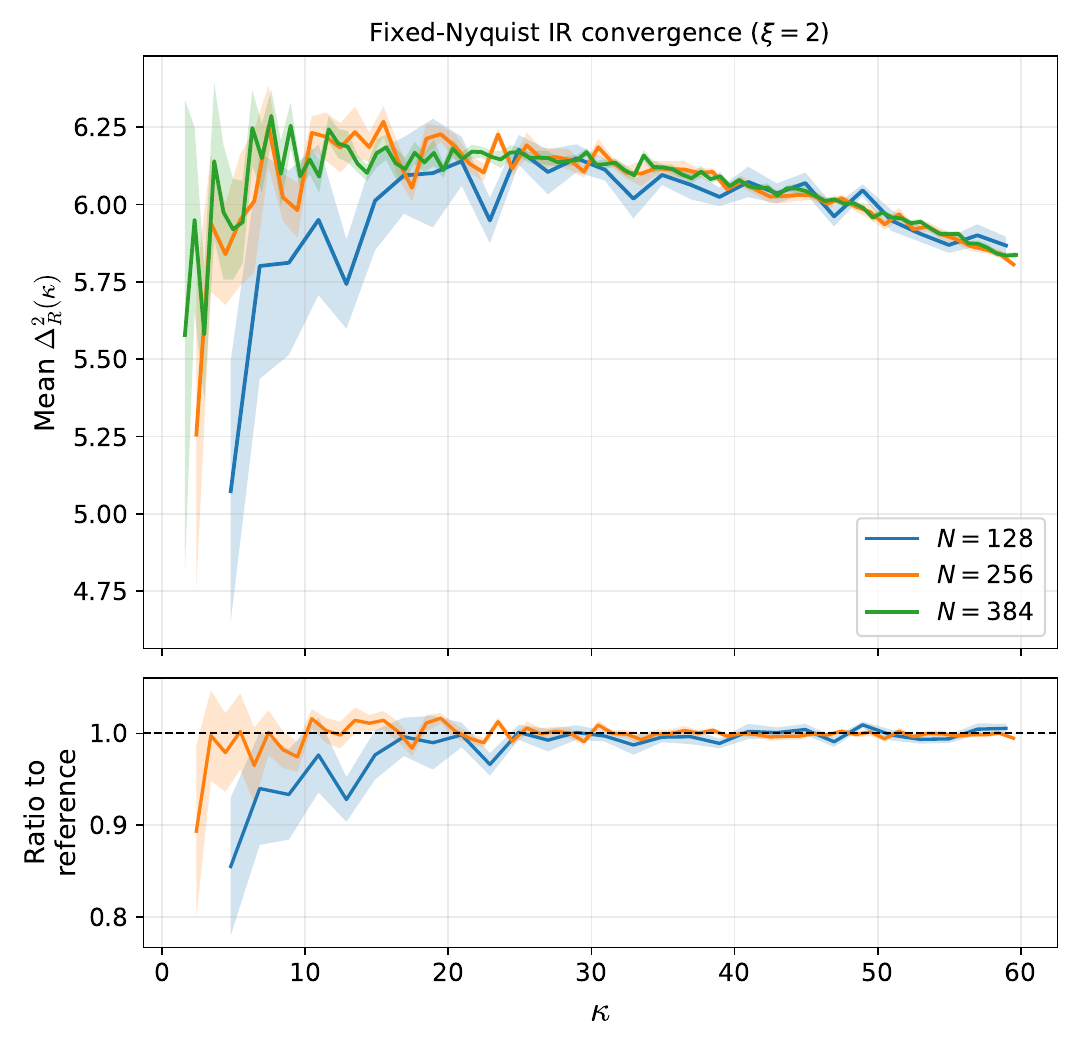}
    \caption{In the top panel, the dimensionless response power spectrum $\Delta^2_{R}(\kappa)$ with $\xi=2$ is shown for the different combinations $(\rm{N}, \tilde{L}) \in \{(128,\pi), (256, 2\pi), (384, 3\pi)\}$, which correspond to the grid resolutions $\Delta \kappa \in \{2, 1, \frac{2}{3}\}$, respectively. In the bottom panel, the ratio of each of the two smaller grid power spectra and the $N = 384$ reference are shown. The $N = 256$ and $N = 384$ curves show good agreement for the entire $\kappa$ range, indicating that integral (\ref{curvature}) has already converged in the IR for $\Delta \kappa = 1$. }
    \label{fig:IR-xi2}
\end{figure}

\begin{figure}[htbp]
    \centering
    \includegraphics[width=0.6\textwidth]{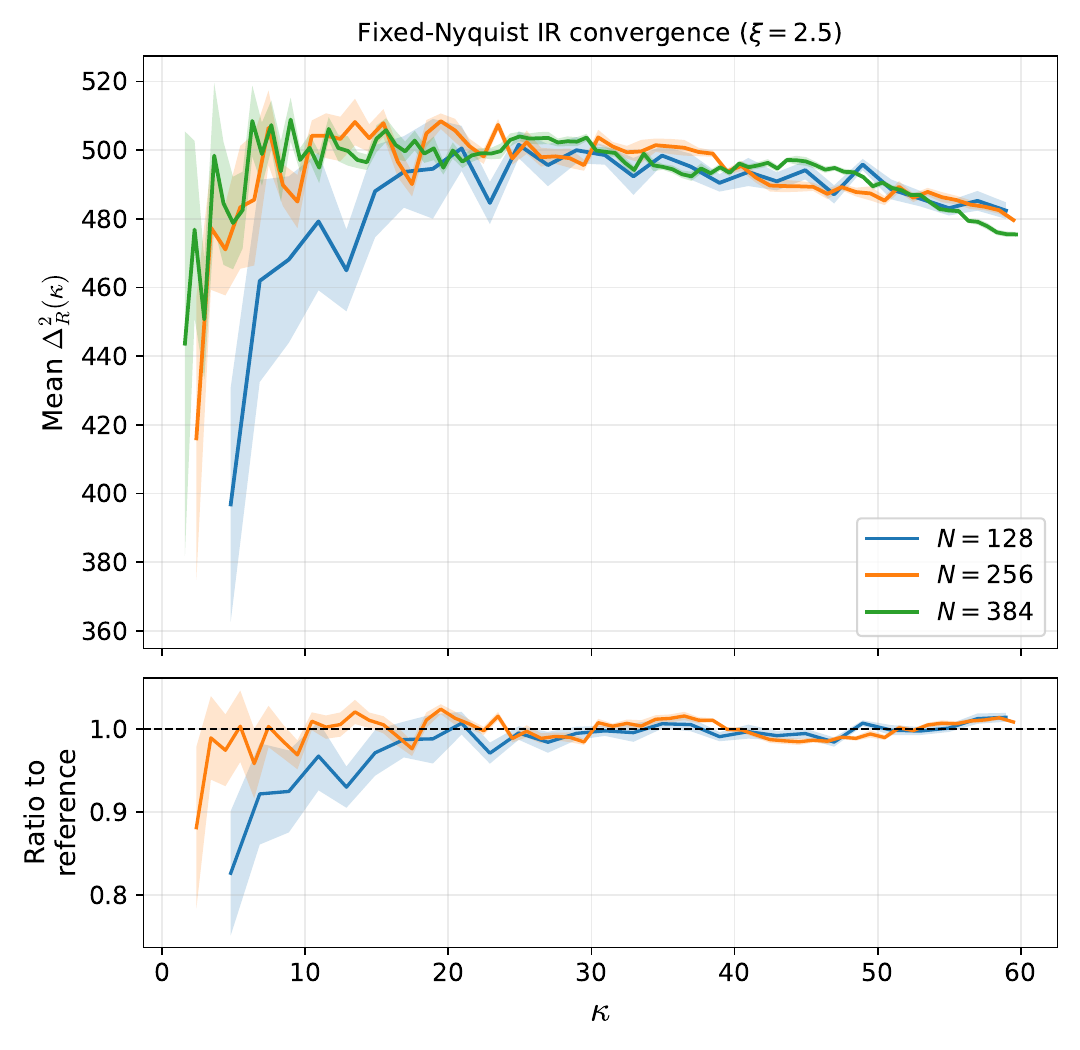}
    \caption{In the top panel, the dimensionless response power spectrum $\Delta^2_{R}(\kappa)$ with $\xi=2.5$ is shown for the different combinations $(\rm{N}, \tilde{L}) \in \{(128,\pi), (256, 2\pi), (384, 3\pi)\}$, which correspond to the grid resolutions $\Delta \kappa \in \{2, 1, \frac{2}{3}\}$, respectively. In the bottom panel, the ratio of each of the two smaller grid power spectra and the $N = 384$ reference are shown.}
    \label{fig:IR-xi2p5}
\end{figure}

\begin{figure}[htbp]
    \centering
    \includegraphics[width=0.6\textwidth]{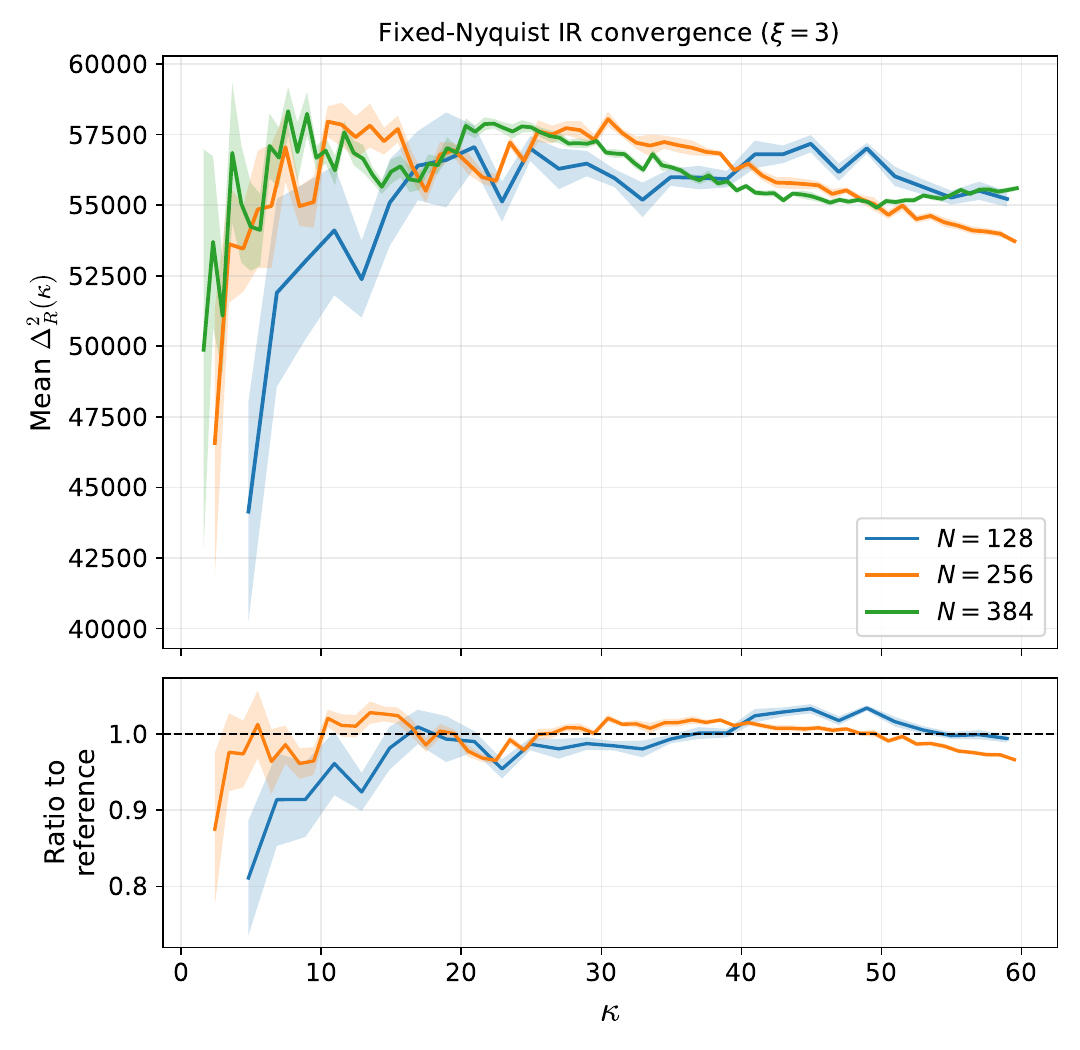}
    \caption{In the top panel, the dimensionless response power spectrum $\Delta^2_{R}(\kappa)$ with $\xi=3$ is shown for the different combinations $(\rm{N}, \tilde{L}) \in \{(128,\pi), (256, 2\pi), (384, 3\pi)\}$, which correspond to the grid resolutions $\Delta \kappa \in \{2, 1, \frac{2}{3}\}$, respectively. In the bottom panel, the ratio of each of the two smaller grid power spectra and the $N = 384$ reference are shown.}
    \label{fig:IR-xi3}
\end{figure}

The dimensionless length of the Fourier grid $\tilde L$ determines the IR bound. To ensure convergence in the IR, we simultaneously increase $\tilde L$ and N by the same factor and measure $\Delta^2_{R}(\kappa)$. This approach holds the ratio $N/\tilde{L}$ constant, leaving $\kappa_{Ny}$ fixed. We test the combinations $(N,\tilde{L}) \in \{(128,\pi), (256, 2\pi), (384, 3\pi)\}$. These corresponds to the grid resolutions $\Delta \kappa \in \{2, 1, \frac{2}{3}\}$, respectively. The results for $\xi \in \{2,2.5,3\}$ are shown in Figures \ref{fig:IR-xi2}, \ref{fig:IR-xi2p5}, and \ref{fig:IR-xi3}. The bottom panels of these figures show the ratio of the smaller-grid power spectra to the $N=384$ reference. Due to the variable spacing of the different resolution grids, it was not possible to compute their power spectra at the same values of $\kappa$ and take their ratio. To circumvent this issue, we linearly interpolated each reference ($N=384$) spectrum and computed their average at the smaller grid $\kappa$ values.  

We see that for small $\kappa$ integral (\ref{curvature}) is not properly resolved, just as in Section \ref{UV_section}. Here the effect is even worse for the coarse grid of resolution $\Delta \kappa = 2$, which only begins to converge with the other two curves around $\kappa \approx 17$. The $N = 256$ curve, however, is in good agreement with the $N=384$ reference for the entire $\kappa$ range, indicating that a resolution of $\Delta \kappa = 1$ is sufficient for IR convergence.    

\section{Validation} \label{Validation}
We now turn to the literature to validate our method against previously published results for the power spectrum and bispectrum. All tests discussed in this section were computed with realizations that were generated with grid size $N_g = 256^3$ and dimensionless box length $\tilde L = 2\pi$. Measurements of the power spectrum and bispectrum were performed for $\xi \in \{2,2.5,3\}$ using \texttt{pylians3}. For the time integration in Eq. (\ref{curvature}), we use a late-time bound $\tilde{\tau}=-0.09$ and number of sampled points $N_t = 100, 28,$ and $14$ for $\xi=2,2.5,$ and $3$, respectively. These parameters were chosen to given below $1\%$ error for the time integral (see Section \ref{Time Convergence}).   

\subsection{Power Spectrum} \label{Power_Spectrum}
We begin with the result for the inverse decay power spectrum given in\footnote{We note that the notation of \cite{Barnaby_2011} differs from our own in that they use the symbol $P_{\zeta}$ for the dimensionless power spectrum, whereas we use $P_{\zeta}$ for the dimensionful power spectrum and $\Delta_{\zeta}^{2 \; \rm inv}$ for the dimensionless power spectrum.} \cite{Barnaby_2011}:
\begin{align}\label{BNP_IDPS}
    \Delta_{\zeta}^{2 \; \rm inv}(k) = \bigg(\frac{k}{k_0}\bigg)^{n_s-1} \mathcal{P}^2 f_{2}(\xi)e^{4\pi \xi},
\end{align}
where $k_0$ is the pivot scale, $\mathcal{P} \equiv (H^2/2\pi|\dot{\bar\phi}|)^2$, and they compute $f_2(\xi)$ by numerically evaluating an integral involving the approximate mode functions (\ref{A_approx}). They provide the monomial fit 
\begin{align}\label{mon_fit}
f_{2}(\xi) \approx \frac{3 \cdot 10^{-5}}{\xi^{5.4}}
\end{align}
for the region $2 \leq \xi \leq 3$. Utilizing Eqs. (\ref{P_inv}) and (\ref{Dimensionless_PS}) to express $\Delta_{\zeta}^{2 \; \rm inv}$ in terms of $\Delta_{R}^{2}$, we have 
\begin{align}\label{f2}
    f_{2}(\xi) = (8\pi^2\xi)^2 \,e^{-4\pi\xi}\, \Delta_{R}^{2}(\kappa). 
\end{align}
We recall that $\Delta_{R}^{2}(\kappa)$ is exactly scale-invariant\footnote{This is true at our level of approximation, assuming constant $\xi$ and de Sitter expansion.}, however our measurement of this quantity will not be, due to approximate numerical convergence and stochastic noise. 

\begin{figure}[H]
    \centering
    \includegraphics[width=0.6\textwidth]{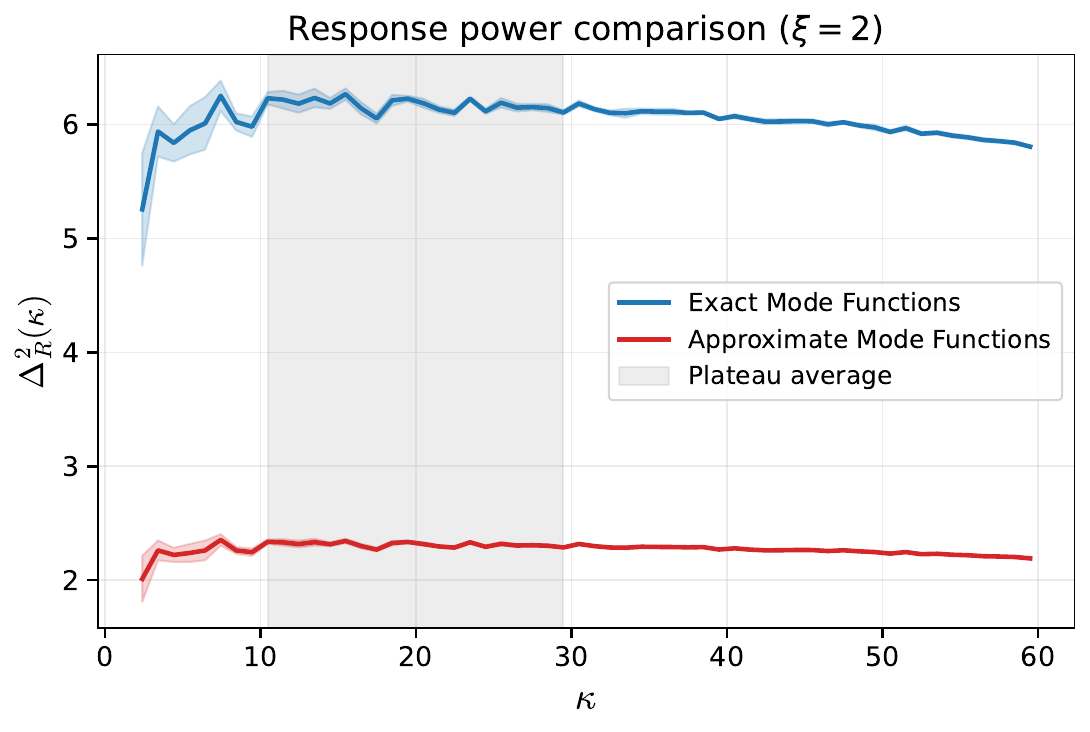}
    \caption{This figure shows the response power spectrum $\Delta^2_{R}(\kappa)$ calculated with the approximate (red) and exact (blue) mode functions for $\xi=2$. The power spectra were computed for each of ten realizations of $R$ and their results were averaged at each $\kappa$. We then average each of the two power spectra over the grey shaded region corresponding to $10 \leq \kappa \leq 30$; the result is reported in Section \ref{Validation}. The colored shaded regions corresponds to the 1$\sigma$ error of the mean.}
    \label{PS_plot_xi=2}
\end{figure}

\begin{figure}[]
    \centering
    \includegraphics[width=0.6\textwidth]{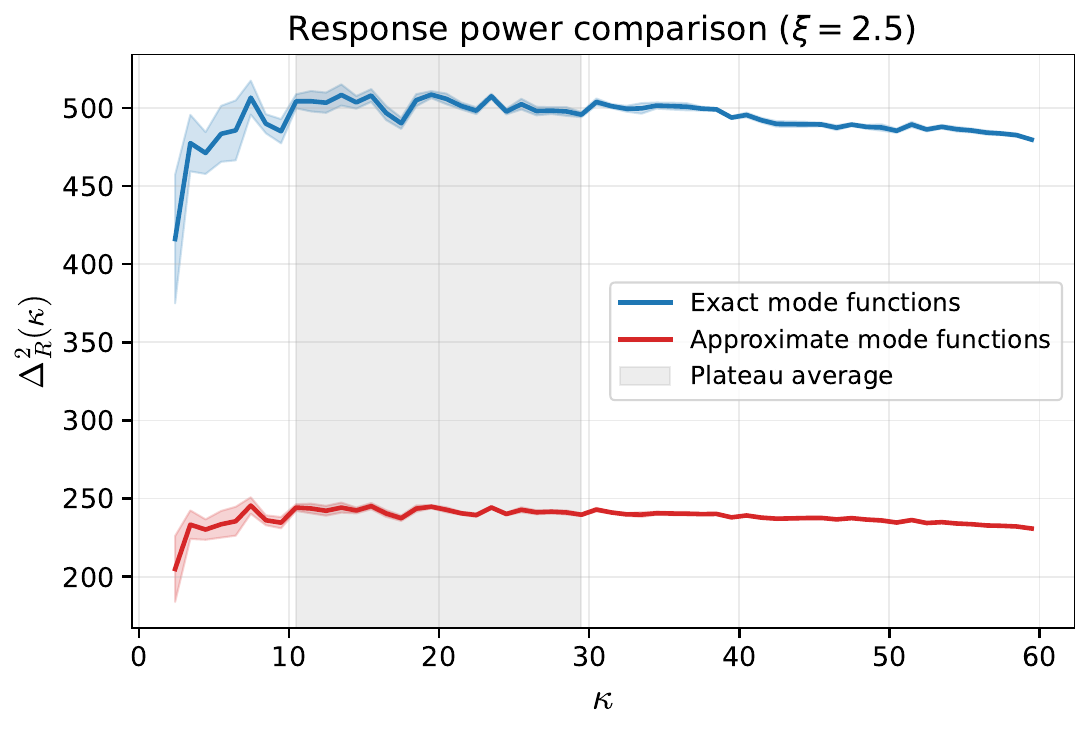}
    \caption{This figure shows the response power spectrum $\Delta^2_{R}(\kappa)$ calculated with the approximate (red) and exact (blue) mode functions for $\xi=2.5$. The colored shaded region corresponds to the 1$\sigma$ error of the mean. See Figure \ref{PS_plot_xi=2} and Section \ref{Validation} for more details.}
    \label{PS_plot_xi=2.5}
\end{figure}

\begin{figure}[]
    \centering
    \includegraphics[width=0.6\textwidth]{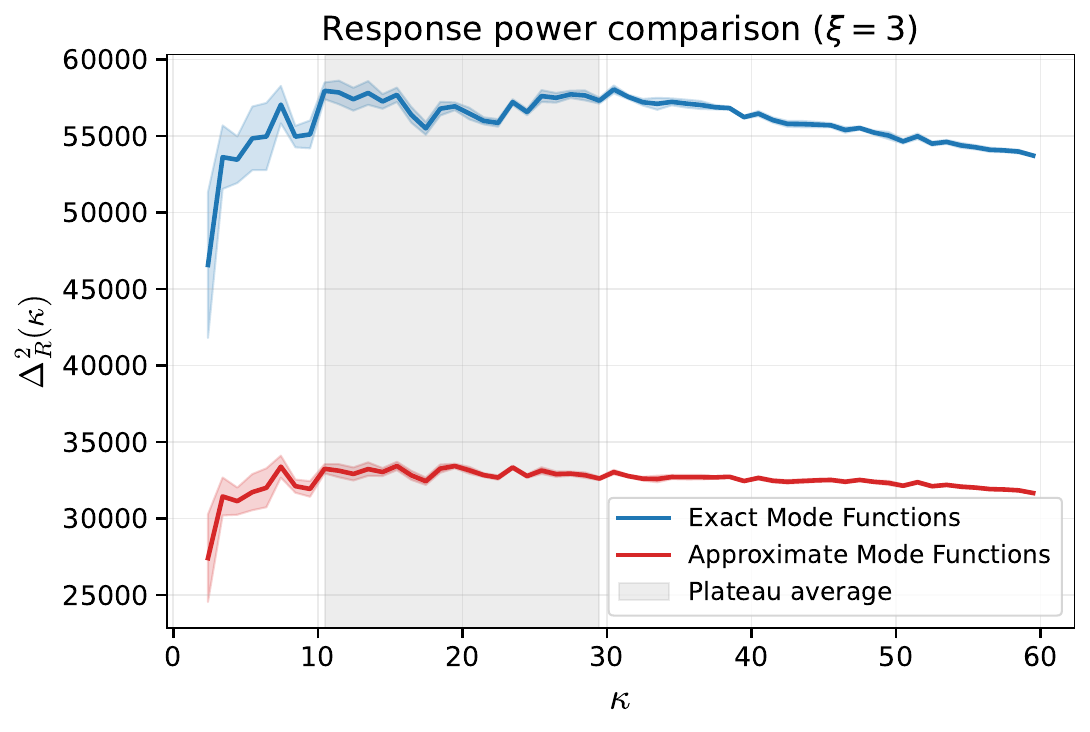}
    \caption{This figure shows the response power spectrum $\Delta^2_{R}(\kappa)$ calculated with the approximate (red) and exact (blue) mode functions for $\xi=3$. The colored shaded region corresponds to the 1$\sigma$ error of the mean. See Figure \ref{PS_plot_xi=2} and Section \ref{Validation} for more details.}
    \label{PS_plot_xi=3}
\end{figure}

To obtain $f_{2}(\xi)$, we first compute the dimensionless power spectrum $\Delta_{R}^{2}(\kappa)$. We generate 10 realizations of the response $R(\vb k)$, compute their power spectra, and average them at each $\kappa$. We performed this procedure twice, once with the exact Coulomb mode functions (\ref{A_m}), and once with the approximate mode functions (\ref{A_approx}) used by \cite{Barnaby_2011}. We find a surprisingly large discrepancy between these two power spectra. These results are shown in Figures \ref{PS_plot_xi=2}, \ref{PS_plot_xi=2.5}, and \ref{PS_plot_xi=3}. We compute the average value of $\Delta_{R}^{2}(\kappa)$ over the plateu-like region $10\leq\kappa\leq30$, where the integral (\ref{curvature}) has converged. Inserting this value into Eq. (\ref{f2}) then yields $f_2(\xi)$.

\begin{table}[htbp]
    \centering
    \small
    \renewcommand{\arraystretch}{1.3}
    \begin{tabular}{|l|c|c|c|}
        \hline
        & $\xi=2$ & $\xi=2.5$ & $\xi=3$ \\
        \hline
        $\xi^6 f^{\rm exact}_2(\xi)$
        & $1.20\times10^{-4}$
        & $1.08\times10^{-4}$
        & $9.90\times10^{-5}$ \\
        \hline
        $\xi^6 f_2^{++}(\xi)$
        & $1.26\times10^{-4}$
        & $1.08\times10^{-4}$
        & $9.85\times10^{-5}$ \\
        \hline
        $\xi^6 f_2^{\mathrm{approx}}(\xi)$
        & $4.49\times10^{-5}$
        & $5.23\times10^{-5}$
        & $5.73\times10^{-5}$ \\
        \hline
        $\xi^6 f_2^{\mathrm{mon}}(\xi)$
        & $4.55\times10^{-5}$
        & $5.20\times10^{-5}$
        & $5.80\times10^{-5}$ \\
        \hline
    \end{tabular}
    \caption{Comparison of $\xi^6 f_2(\xi)$ calculated via four different methods. $f^{\rm exact}_2$ was computed from realizations of the response $R$, using the exact Coulomb mode functions as described in Section \ref{Validation}. $f_2^{++}$ was extracted from the plots of \cite{Durrer_2024} using the PlotDigitizer tool, $f_2^{\rm approx}$ was computed from realizations of the response $R$ using the approximate mode functions (\ref{A_approx}), and $f_2^{\rm mon}$ was computed from the monomial fit (\ref{mon_fit}).}
    \label{tab:f2_comparison}
\end{table}

\cite{Durrer_2024} provides a more general computation of the power spectrum which includes both helicity modes and metric perturbations. They also separately report their results including only the tachyonically-enhanced modes in one case, and both helicities modes in the other.\footnote{Due to a different Fourier convention, they find that the $A_+(\tau,k)$ modes are amplified.} We use the PlotDigitizer tool\footnote{https://plotdigitizer.com/} to extract numerical values for $\xi^6 f(\xi)$ from the plot of \cite{Durrer_2024} that includes only the enhanced modes. In Table \ref{tab:f2_comparison}, we compare our results with those of \cite{Durrer_2024} for the tested values of $\xi \in \{2,2.5,3\}$ and find excellent agreement. The table also includes our results obtained from using the approximate mode functions (\ref{A_approx}) to generate the response $R$, as well as values tabulated via the monomial fit (\ref{mon_fit}). 

We note that our method of averaging the power spectra of the stochastic response $R$ to obtain $f_2(\xi)$ yields results that match the fit (\ref{mon_fit}) provided by \cite{Barnaby_2011} extremely well when using their same approximate mode functions. However, performing the same measurement with the exact Coulomb mode function gives a significantly different results (as large as a factor of $\sim2.5$ difference for the case of $\xi=2$).  

\subsection{Bispectrum} \label{Bispectrum}
We next turn to the inverse decay bispectrum $B_{\zeta}^{\rm inv}$, which is defined by
\begin{align}\label{bispectrum_def}
   \expval{\zeta^{\rm inv}(\vb k_1) \zeta^{\rm inv}(\vb k_2) \zeta^{\rm inv}(\vb k_3)} \equiv (2\pi)^3 B_{\zeta}^{\rm inv}(k_1, k_2, k_3)\, \Dird(\mathbf{k}_1+\mathbf{k}_2+\mathbf{k}_3),
\end{align}
where it is understood that we have evaluated the correlator at some time towards the end of inflation.
An expression for the inverse decay bispectrum was also given in\footnote{We have adjusted the factors of $2\pi$ in this expression to account for both our different Fourier convention and the fact that \cite{Barnaby_2011} defines the bispectrum to include the factor of $(2\pi)^3$ in (\ref{bispectrum_def}).} \cite{Barnaby_2011}:
\vspace{5mm}
\begin{align} \label{Bispectrum_eqn}
B_{\zeta}^{\rm inv}(k_1, k_2, k_3) = 
\frac{3}{10}(2\pi)^{4} \, \mathcal{P}^3 e^{6\pi\xi}
\frac{1+x_2^3+x_3^3}{k^6 x_2^3 x_3^3}
f_3(\xi,x_2,x_3), 
\end{align}
where $|\vb k_1| = k$, $|\vb k_2| = x_2 \, k$, $|\vb k_3| = x_3 \, k$, and  $f_3(\xi,x_2,x_3)$ is a high-dimensional integral which must be computed numerically. \cite{Barnaby_2011} evaluated this integral using the approximate mode function (\ref{A_approx}) for the equilateral case $x_2=x_3=1$, and provide the monomial fit:
\begin{align} \label{f3_fit}
    f_3(\xi, 1, 1) \approx \frac{7.4 \cdot 10^{-8}}{\xi^{8.1}} \hspace{7mm} \text{for} \hspace{5mm} 2\leq \xi \leq 3.  
\end{align}

The relationship between the bispectra of the inverse decay curvature perturbation $\zeta^{\rm inv}$ and the response $R$ is
\begin{align} \label{Bispectra_Rtozeta}
    B_R(k_1, k_2, k_3) = \bigg[8\pi^2 \xi\, \mathcal{P} \bigg]^{-3} B_{\zeta}^{\rm inv}(k_1, k_2, k_3),
\end{align}
where we have neglected the mild scale dependence as done in \cite{Barnaby_2011}. We define the dimensionless bispectrum 
\begin{align}\label{BR_to_dimensionless}
\tilde{B}_R(k_1, k_2, k_3) \equiv \frac{(k_1k_2k_3)^2}{(2\pi^2)^2} B_R(k_1, k_2, k_3).    
\end{align}
From Eqs. (\ref{Bispectrum_eqn}), (\ref{Bispectra_Rtozeta}), and (\ref{BR_to_dimensionless}), we find
\begin{align}\label{f3_BR}
    f_3(\xi, 1, 1) = \frac{1280}{9} \pi^6 \xi^3 e^{-6\pi\xi} \tilde{B}_R(\kappa , \kappa, \kappa),
\end{align}
where we have switched to the dimensionless coordinate $\kappa$.

\begin{figure}[H]
    \centering
    \includegraphics[width=0.6\textwidth]{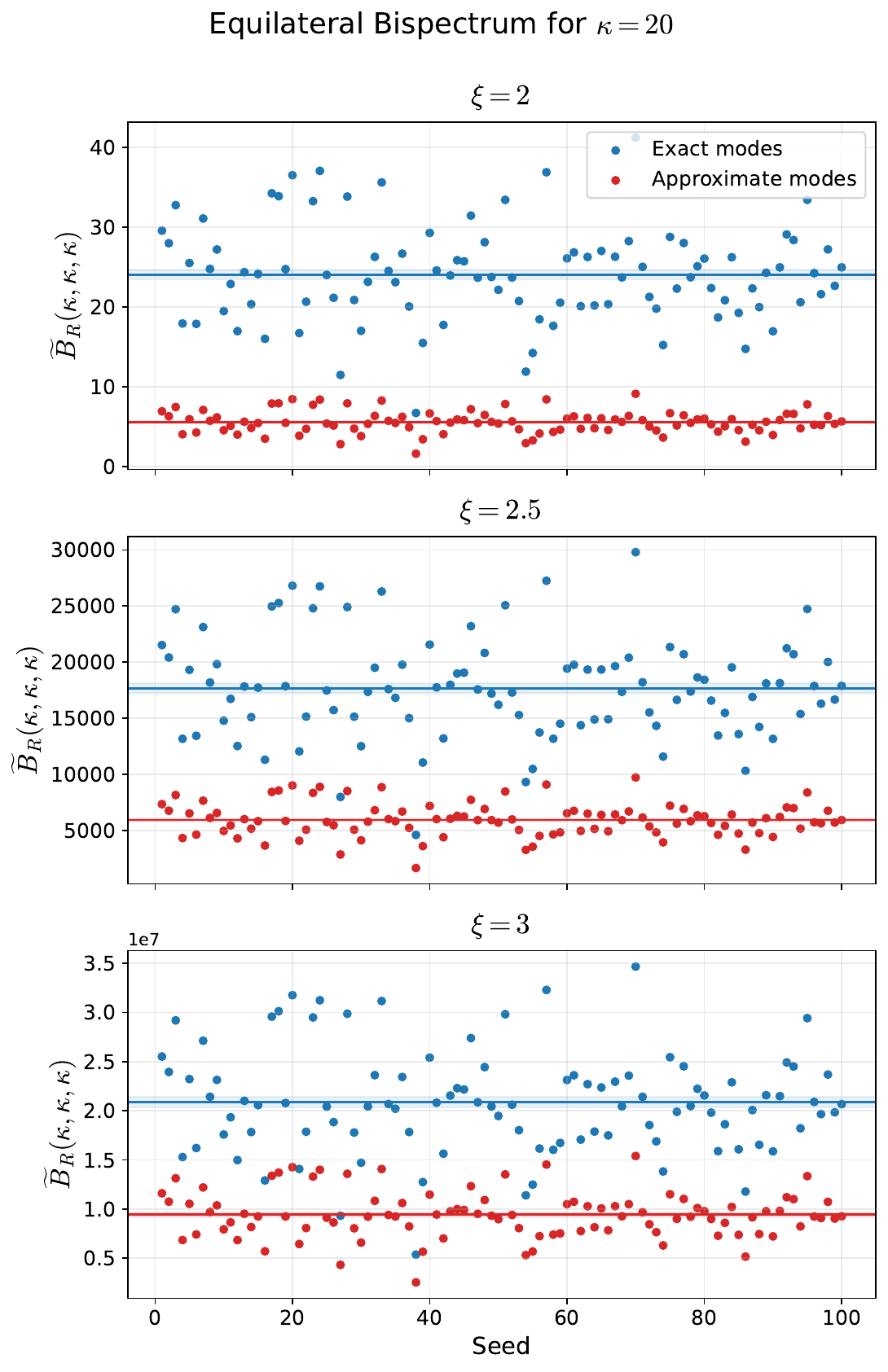}
    \caption{The bispectrum $\tilde{B}_R(\vb \kappa, \vb \kappa, \vb \kappa)$ measured from 100 realizations of the response function is shown for $\xi=2,2.5,$ and $3$ using both the exact Coulomb mode functions (blue) and the approximate mode functions (red). Each dot represents $\tilde{B}_R(\kappa , \kappa, \kappa)$ for a single realization corresponding to the random seed shown on the x-axis. The sample mean is shown as a horizontal line with the shaded region representing the standard error of the mean.}
    \label{Bispectrum_plot}
\end{figure}

We generate 100 realizations of the response $R$ and measure the dimensionless bispectrum for each realization in the equilateral configuration $\kappa = 20, x_2 = 1, x_3 = 1$. We then take the average of these measurements. We choose the value $\kappa = 20$ since this sits in the plateu-like region of the power spectrum (see Figure \ref{PS_plot_xi=2}), where the integral (\ref{curvature}) has converged. We perform this procedure for both the exact mode functions (\ref{A_m}) and the approximate mode functions (\ref{A_approx}) for $\xi=2,2.5,$ and $3$. The bispectrum value corresponding to each random realization and their average is shown in Figure \ref{Bispectrum_plot}.

After computing $\tilde{B}_R(\kappa , \kappa, \kappa)$ for different values of $\xi$, we insert these results into Eq. (\ref{f3_BR}) to obtain $f_3(\xi, 1, 1)$; the results are shown in Table \ref{tab:f3_comparison}. We again find good agreement between our results and those of \cite{Durrer_2024}, with a maximum difference of $\sim$$4\%$ for the case of $\xi=2$. This difference falls below $1\%$ for case of $\xi=3$; this could be due to the increasing ease of numerical convergence of the time integral in Eq. (\ref{curvature}) as one moves to larger $\xi$, as discussed in Section \ref{Time Convergence}. 

Just as in Section \ref{Power_Spectrum}, we find that measurements performed with the approximate mode functions differ appreciably from those performed with the exact mode functions. Here too the approximate results $f_3^{\rm approx}$ agree quite well with the monomial fit (\ref{f3_fit}) provided by \cite{Barnaby_2011}, but these two results differ by, at worst, a factor of $\sim$$4$ (for the case of $\xi=2$) from the exact mode results.   

\begin{table}[htbp]
    \centering
    \small
    \renewcommand{\arraystretch}{1.3}
    \begin{tabular}{|l|c|c|c|}
        \hline
        & $\xi=2$ & $\xi=2.5$ & $\xi=3$ \\
        \hline
        $\xi^9 f_3^{\mathrm{exact}}(\xi)$
        & $5.71\times10^{-7}$
        & $4.93\times10^{-7}$
        & $4.19\times10^{-7}$ \\
        \hline
        $\xi^9 \left|f_3^{+++}(\xi)\right|$
        & $5.95\times10^{-7}$
        & $4.83\times10^{-7}$
        & $4.16\times10^{-7}$ \\
        \hline
        $\xi^9 f_3^{\mathrm{approx}}(\xi)$
        & $1.32\times10^{-7}$
        & $1.66\times10^{-7}$
        & $1.90\times10^{-7}$ \\
        \hline
        $\xi^9 f_3^{\mathrm{mon}}(\xi)$
        & $1.38\times10^{-7}$
        & $1.69\times10^{-7}$
        & $1.99\times10^{-7}$ \\
        \hline
    \end{tabular}
    \caption{Comparison of $\xi^9 f_3(\xi)$ obtained via four different methods. $f_3^{\rm exact}$ and $f_3^{\rm approx}$ were computed from realizations of the stochastic response $R$ using the exact and approximate mode functions, respectively, as described in Section \ref{Validation}. $|f_3^{+++}|$ was extracted from the plots of \cite{Durrer_2024} with the PlotDigitizer tool, and $f_3^{\rm mon}$ was tabulated using the monomial fit (\ref{f3_fit}).}
    \label{tab:f3_comparison}
\end{table}

\section{Conclusion} \label{Section6}
We have developed a new method for efficiently computing inverse decay polyspectra for axion inflation. The fact that the inverse decay inflaton perturbations are sourced by an approximately classical gauge field allowed us to generate stochastic realizations of these fields with correlators that match the corresponding quantum correlators. We presented this method: detailing our implementation of a 3/2 dealiasing scheme, Legendre-Gauss integration, and a UV cutoff. Furthermore, we tested different values of the Nyquist wavenumber, grid resolution, and sampling rate to ensure convergence of each integral. 

We computed the power spectrum and bispectrum, and found a significant discrepancy between our results and those of \cite{Barnaby_2011}. This discrepancy appears to be due to \cite{Barnaby_2011} using approximation (\ref{A_approx}) for the gauge field mode functions, as it disappeared when we used these same mode functions in our code. Our results for the power spectrum and bispectrum are in excellent agreement with \cite{Durrer_2024} for all tested values of $\xi$.  

The Chern-Simons interaction between the axion and the gauge field dynamically breaks parity symmetry, with the parity-odd signal first appearing in the one-loop contribution to the trispectrum. While previous works have analytically computed the trispectrum for a few different configurations of the wavevectors \cite{Niu_2023, Fujita_2024}, much of the parameter space remains unexplored. With this new method one can efficiently generate a large number of curvature perturbation realizations and use them to compute the trispectrum (or generally any polyspectrum) numerically. Such a tool will be highly useful in searching for the signal of this inflationary model in LSS data. We leave the computation of the primordial trispectrum to future work.

In addition to allowing for efficient computation of primordial correlators, the realizations of the curvature perturbations can provide initial conditions for LSS simulations \cite{Springel_2021}, and cosmological perturbation theory codes such as GridSPT \cite{Taruya_2018}. With these one can obtain the matter density field and---after applying a galaxy biasing scheme and accounting for redshift-space distortions---the galaxy density field. 

Correct estimation of the covariance matrix in previous measurements of the galaxy 4PCF has proven difficult due to the very large number of degrees of freedom \cite{Philcox_2022, Hou_2023, Krolewski_2024, slepian2025, Hou_2026}. In searching for the signal of a particular model, such as axion inflation, we can rank the tetrahedron configurations by their signal-to-noise ratio (SNR) and include only the highest SNR configurations in our analysis. This will significantly improve the estimate of the covariance matrix, leading to a more robust measurement of the 4PCF.

\appendix
\section{Nyquist Modes}\label{AppendixA}
In Section \ref{Section3} we asserted that all Fourier modes of the gauge field must vanish when evaluated at grid points where at least one of the wave vector's components is equal to the Nyquist wavenumber. Here we provide a proof of this claim. Consider the wave vector $\vb k = (k_{Ny}, k_x, k_z)$, where without loss of generality, we have assumed the x-component to be the Nyquist wavenumber, and we will assume the other components are nonzero. Every grid point corresponds to some vector $\vb n$, such that
\begin{align}
    \vb q = \frac{2\pi}{L} \vb n,
\end{align}
for an arbitrary wave vector $\vb q$. The components $n^i$ each have a negative partner on the grid, $-n^i$, except for Nyquist components of $n_{Ny}^i = N/2$. The Nyquist components and their negative values correspond to the same point on the Fourier grid. 

Let's now consider the random realizations $\vb X_{-}(\vb k)$ which were defined to satisfy the polarization vector identities (\ref{pol_vec}). From transversality, we have:
\begin{align}\label{kdotX1}
    \vb k \vdot \vb X_-(\vb k) = 0 \implies k_{Ny}f + k_yg + k_z h = 0,
\end{align}
where $f, g,$ and $h$ are the complex components of the vector $\vb X_-(\vb k)$. Since the Nyquist component and its negative correspond to the same grid point,
\begin{align}
    \vb X_-(k_{Ny}, k_y, k_z) = \vb X_-(-k_{Ny}, k_y, k_z). 
\end{align}
Therefore, we also have
\begin{align}\label{kdotX2}
     (-k_{Ny}, k_y, k_z) \vdot \vb X_-(-k_{Ny}, k_y, k_z) = 0 \implies -k_{Ny}f + k_yg + k_z h = 0.
\end{align}
From Eqns. (\ref{kdotX1}) and (\ref{kdotX2}), we find 
\begin{align}\label{trans_condition}
    f = 0 \hspace{10mm} \text{and} \hspace{10mm} g = -\frac{k_z}{k_y}h.
\end{align}
Using the cross product property of (\ref{pol_vec}), we have
\begin{align}\label{cross_identity}
    -i\vb k \cross \vb X_{-}(\vb k) = k \vb X_-(\vb k),
\end{align}
where $k = \sqrt{k_{Ny}^2 + k_y^2 + k_z^2}$ is the magnitude of the wavevector. Let us consider just the y-component of Eqn. (\ref{cross_identity}):
\begin{align} \label{y-component}
    ik_{Ny}h = kg.
\end{align}
In Eq. (\ref{y-component}), $h$ and $g$ are related by an imaginary proportionality constant, whereas in Eq. (\ref{trans_condition}), these two numbers are related by a real number. This is only possible if $h=g=0$. We have thus found that given a wave vector $\vb k$ with at least one component equal to the Nyquist wavenumber, we must have 
\begin{align}
    \vb X_-(\vb k) = 0.
\end{align}

\section{The Classical Field Approximation}\label{AppendixB}
Throughout this work we assume that the inflaton and gauge fields are approximately classical, i.e. that their quantum correlators are approximately equal to the ensemble average of correlators computed with randomly generated c-number valued fields. A quantum correlator of multiple copies of a field operator, all evaluated at the same time, exactly equals the corresponding classical correlator \cite{Ashtekar_2020}. Unequal time correlators, however, have an irreducibly quantum piece which cannot be reproduced by c-number fields. 

When computing the vacuum power spectrum:
\begin{align}
    \expval{\zeta^{\rm vac}(\tau, \vb k_1) \zeta^{\rm vac}(\tau, \vb k_2)} = (2\pi)^3 P^{\rm vac}_{\zeta}(k) \Dird(\vb k_1 + \vb k_2),
\end{align}
the curvature perturbation operators are evaluated at the same time. We may therefore obtain this quantum correlator by taking the ensemble average of power spectra computed from classical field configurations generated from the probability distribution corresponding to the quantum state; in this case a Gaussian. 

For the inverse decay power spectrum, the situation is complicated by the fact that the curvature perturbations $\zeta^{\rm inv}$ are sourced by the gauge field through Eq. (\ref{curvature}). This leads to unequal-time correlators of the gauge field and its derivative in Fourier space. 
 We will find that the classical and quantum correlators involving $\zeta^{\rm inv}$ are approximately equal if the imaginary part of the gauge field mode functions is significantly smaller than the real part, or equivalently, the commutator of the fields at unequal times is very small when compared to their anticommutator.

The Fourier Transform of the gauge field is
\begin{align}\label{A_FT}
\tilde{A}^{\,i}(\tau,\vb{q})
=
\hat{\epsilon}^{\,i}_{-}(\vb{q})
\left[
a_{-}(\vb{q})A_-(\tau,q)
+
a_{-}^{\dagger}(-\vb{q})A^{*}_-(\tau,q)
\right].
\end{align}
Here, we are not yet assuming the mode functions $A_-(\tau,q)$ are real as was done in the main text. Since the inverse decay axion perturbations are sourced by the gauge field, any correlator of the inverse decay perturbations involves a correlator of the source $J$, which was defined in Eq. (\ref{J_def}). Schematically,
\begin{align}
    \expval{\zeta^{\rm inv}(\tau, \vb k_1) \zeta^{\rm inv}(\tau, \vb k_2) \cdots} \propto \expval{J(\tau_1, \vb{k_1}) J(\tau_2, \vb{k_2}) \cdots },
\end{align}

where the source functions are evaluated at different times since each is integrated against a Green's function as was shown in Eq. (\ref{inv_mf}). From the definition of the source J given in (\ref{J_def}) and (\ref{A_FT}), we have
\begin{align}
J(\tau, \vb{k}) = \frac{i}{a(\tau)\Lambda} \epsilon^{lmn}
\int \frac{\dd^3\vb{q}}{(2\pi)^3}\,q_m \tilde{A}'_l(\tau, \vb k - \vb q) \tilde{A}_n(\tau, \vb q),
\end{align}
where $\epsilon^{lmn}$ is the Levi-Civita symbol and $l,m,$ and $n$ denote spatial indices.

We now consider the unequal-time source power spectrum:
\begin{align}
\expval{J(\tau_1, \vb{k_1}) J(\tau_2, \vb{k_2})} &= -\frac{\epsilon^{ijk}\epsilon^{lmn}}{a(\tau_1)a(\tau_2)\Lambda^2} \int \frac{\dd^3\vb{q_1}}{(2\pi)^3} \int \frac{\dd^3\vb{q_2}}{(2\pi)^3}\,q_{1j} \,q_{2m} \nonumber  \\
& \times  \bigg < \tilde{A}'_i(\tau_1, \vb k_1 - \vb q_1) \tilde{A}_k(\tau_1, \vb q_1) \tilde{A}'_l(\tau_2, \vb k_2 - \vb q_2) \tilde{A}_n(\tau_2, \vb q_2) \bigg >.
\end{align}
Applying Wick's theorem, we have
\begin{align} \label{Wick_Thm}
    &\bigg < \tilde{A}'_i(\tau_1, \vb k_1 - \vb q_1) \tilde{A}_k(\tau_1, \vb q_1) \tilde{A}'_l(\tau_2, \vb k_2 - \vb q_2) \tilde{A}_n(\tau_2, \vb q_2) \bigg > \nonumber \\
    & = \big < \tilde{A}'_i(\tau_1, \vb k_1 - \vb q_1) \tilde{A}'_l(\tau_2, \vb k_2 - \vb q_2) \big > \big <\tilde{A}_k(\tau_1, \vb q_1) \tilde{A}_n(\tau_2, \vb q_2) \big> \nonumber \\
    & + \big < \tilde{A}'_i(\tau_1, \vb k_1 - \vb q_1) \tilde{A}_n(\tau_2, \vb q_2) \big > \big < \tilde{A}_k(\tau_1, \vb q_1) \tilde{A}'_l(\tau_2, \vb k_2 - \vb q_2) \big >,
\end{align}
where we have excluded disconnected diagrams. Each 2PCF in (\ref{Wick_Thm}) is the same save for up to two derivatives with respect to conformal time (and different arguments). Let's now focus on the term without any time derivative: 
\begin{align}\label{2pt-A}
    \big <\tilde{A}_k(\tau_1, \vb q_1) \tilde{A}_n(\tau_2, \vb q_2) \big>.
\end{align}
The other terms follow directly from the result for this one. For any two operators U and V,
\begin{align}
    UV = \frac{1}{2}[U,V] + \frac{1}{2}\{U,V\}.
\end{align}
For the expectation values of the commutator and anticommutator, we have
\begin{align} \label{commutator}
    \big< [\tilde{A}_k(\tau_1, \vb q_1), \tilde{A}_n(\tau_2, \vb q_2)] \big> = 2i (2\pi)^3 \hat{\epsilon}^{\,k}_{-}(\vu{q}_1) \hat{\epsilon}^{\,n}_{-}(\vu{q}_2)\Im[A_{-}(\tau_1,q_1) A^*_{-}(\tau_2, q_2)]\Dird(\vb q_1 + \vb q_2),
\end{align}
and
\begin{align} \label{anitcommutator}
    \big< \{\tilde{A}_k(\tau_1, \vb q_1), \tilde{A}_n(\tau_2, \vb q_2)\} \big> = 2(2\pi)^3 \hat{\epsilon}^{\,k}_{-}(\vu{q}_1) \hat{\epsilon}^{\,n}_{-}(\vu{q}_2)\Re[A_{-}(\tau_1,q_1) A_{-}^*(\tau_2, q_2)]\Dird(\vb q_1 + \vb q_2).
\end{align}
From Eqs. (\ref{commutator}) and (\ref{anitcommutator}), we can immediately obtain the result for every other 2PCF in (\ref{Wick_Thm}) by simply taking the appropriate derivatives with respect to conformal time and interchanging the arguments. Due to Wick's theorem, any higher order correlator is constructed from these 2PCFs.  

\begin{figure}[ht]
    \centering
    \includegraphics[width=0.8\textwidth]{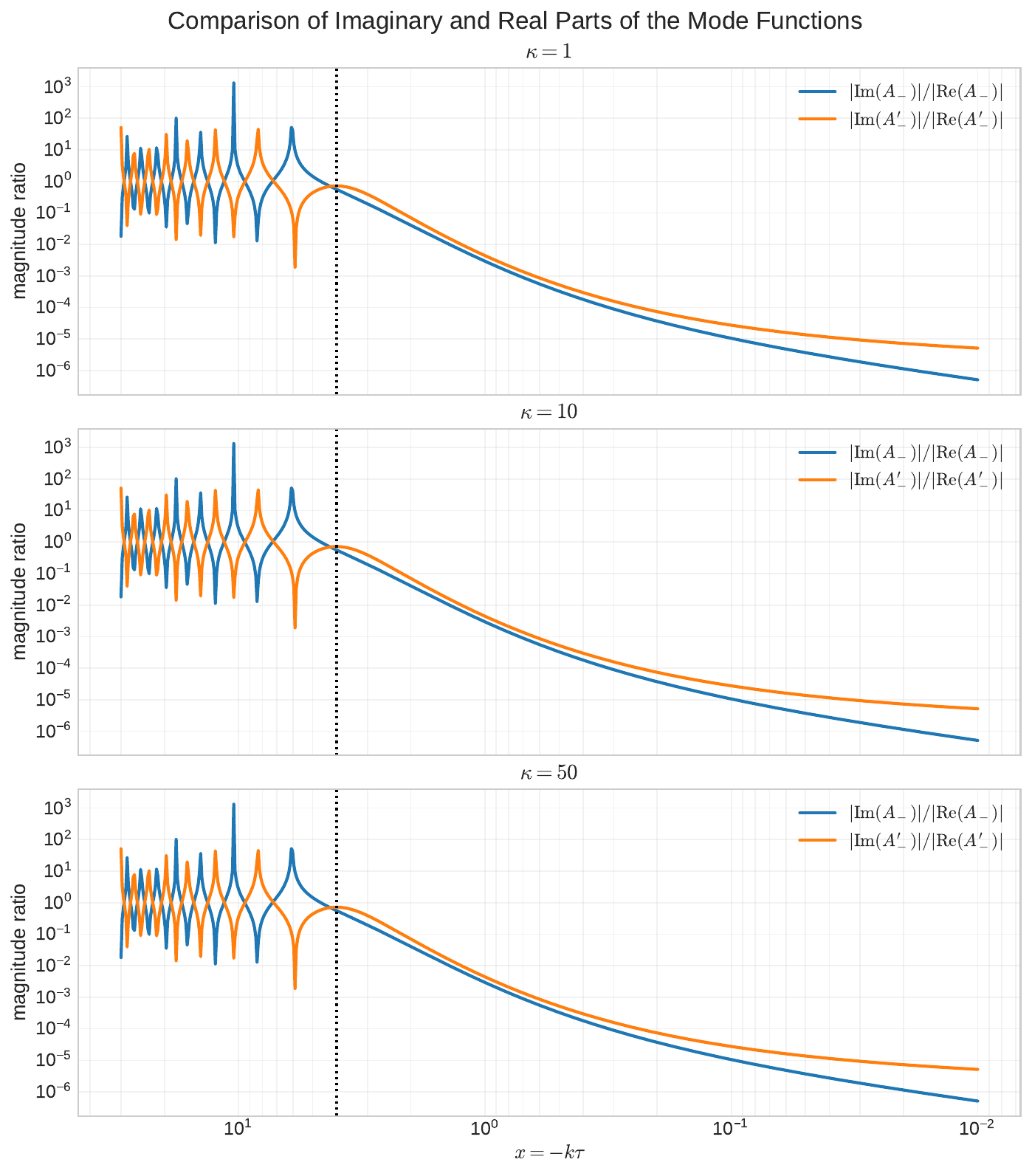}
    \caption{The ratio of the imaginary to the real part of the gauge field mode functions is shown as well as the ratio of their time derivatives. The vertical dotted line corresponds to $-k\tau=2\xi$. The two ratios are $O(1)$ when tachyonic enhancement begins at $-k\tau=2\xi$ and rapidly decrease as time progresses.}
    \label{real_vs_imaginary}
\end{figure}

If the imaginary part of the mode functions $A_{-}(\tau,q)$ are significantly smaller than their real part, then the commutator contribution to the correlator is negligible when compared to the contribution from the anticommutator. In Figure \ref{real_vs_imaginary}, we plot the ratio of the real and imaginary parts of the gauge field mode function and its time derivative, showing that the imaginary part is in fact negligible when compared to the real part for most of the parameter space.

Let us now illustrate how the classical approach reproduces the anti-commutator piece of (\ref{2pt-A}). Consider a classical field, discretized on a cubic lattice of length $L$, and defined by 
\begin{align}\label{A_Classical}
\tilde{A}_{class}^{\,i}(\tau,\vb{q}) 
=
\frac{1}{\sqrt{2}}\hat{\epsilon}^{\,i}_{-}(\vb{q})
\left[
Y(\vb q)A_-(\tau,q)
+
Y^*(-\vb{q})A^{*}_-(\tau,q)
\right],
\end{align}
with $Y(\vb q)$ being a complex-valued Gaussian random field satisfying
\begin{align}
    \expval{Y} = 0, \hspace{5mm} \expval{Y(\vb q)Y(\vb q')} = 0, \hspace{5mm} \text{and} \hspace{5mm} \expval{Y(\vb q)Y^*(\vb q')} = L^3 \kron_{\vb q \,\vb q'}.
\end{align}
Then for the classical correlator, we have
\begin{align}\label{classical_correlator}
    \big <\tilde{A}_{class}^k(\tau_1, \vb q_1) \tilde{A}_{class}^n(\tau_2, \vb q_2) \big> =  L^3 \kron_{\vb q_1  -\vb q_2} \hat{\epsilon}^{\,k}_{-}(\vb{q_1}) \hat{\epsilon}^{\,n}_{-}(\vb{q_2}) \Re [A_{-}(\tau_1,q_1) A_{-}^*(\tau_2, q_2)].
\end{align}
Upon discretizing the Dirac delta function in Eq. (\ref{anitcommutator}), we have 
\begin{align}
     (2\pi)^3\Dird(\vb q_1 + \vb q_2) \to L^3 \kron_{\vb q_1 -\vb q_2}. 
\end{align}
Therefore in our numerical implementation, the classical correlator (\ref{classical_correlator}) exactly matches the anticommutator piece of the quantum correlator. Indeed, the anitcommutator is automatically Weyl ordered, and therefore its expectation value should exactly match the corresponding classical expectation value \cite{Ashtekar_2020}.  The commutator piece of the correlator is irreducibly quantum and so we could not possibly reproduce Eq. (\ref{commutator}) with this classical approach. This is to be expected as classical fields commute, causing this term to vanish.

\acknowledgments
MR acknowledges funding as an International Research Fellow of Japan Society for the Promotion of Science, which made this work possible. The idea for this work was conceived at the Yukawa Institute for Theoretical Physics (YITP) at Kyoto University, and MR is very grateful to Atsushi Taruya for hosting him there. AT acknowledges support from the FY2025 MUSUBIME program of Kyoto University and JSPS KAKENHI Grant Numbers JP23K25868 and JP26H02044. MR thanks Giorgia Biselli, Jason Kristiano, Masahiro Takada, Tomohiro Fujita, and Regan Bhatta for helpful discussions. The authors acknowledge UFIT Research Computing for providing computational resources on HiPerGator and support that have contributed to the research results reported in this publication. We acknowledge the use of ChatGPT 5.5 and 5.6 for code development, but not for the writing of this manuscript.

\bibliographystyle{utphys}
\bibliography{bibliography.bib}

@misc{reinhard2024,
      title={Full Parity-Violating Trispectrum in Axion Inflation: Reduction to Low-D Integrals}, 
      author={Matthew Reinhard and Zachary Slepian and Jiamin Hou and Alessandro Greco},
      year={2024},
      eprint={2412.16037},
      archivePrefix={arXiv},
      primaryClass={astro-ph.CO},
      url={https://arxiv.org/abs/2412.16037}
}

@article{Philcox_2022,
   title={Probing parity violation with the four-point correlation function of BOSS galaxies},
   volume={106},
   ISSN={2470-0029},
   url={http://dx.doi.org/10.1103/PhysRevD.106.063501},
   DOI={10.1103/physrevd.106.063501},
   number={6},
   journal={Physical Review D},
   publisher={American Physical Society (APS)},
   author={Philcox, Oliver H. E.},
   year={2022},
   month=sep }

@article{Barnaby_2011,
   title={Phenomenology of a pseudo-scalar inflaton: naturally large nongaussianity},
   volume={2011},
   ISSN={1475-7516},
   url={http://dx.doi.org/10.1088/1475-7516/2011/04/009},
   DOI={10.1088/1475-7516/2011/04/009},
   number={04},
   journal={Journal of Cosmology and Astroparticle Physics},
   publisher={IOP Publishing},
   author={Barnaby, Neil and Namba, Ryo and Peloso, Marco},
   year={2011},
   month=Apr, pages={009–009} }

@article{Anber_2010,
   title={Naturally inflating on steep potentials through electromagnetic dissipation},
   volume={81},
   ISSN={1550-2368},
   url={http://dx.doi.org/10.1103/PhysRevD.81.043534},
   DOI={10.1103/physrevd.81.043534},
   number={4},
   journal={Physical Review D},
   publisher={American Physical Society (APS)},
   author={Anber, Mohamed M. and Sorbo, Lorenzo},
   year={2010},
   month=Feb }

@article{LargeNG,
   title={Large Non-Gaussianity in Axion Inflation},
   volume={106},
   ISSN={1079-7114},
   url={http://dx.doi.org/10.1103/PhysRevLett.106.181301},
   DOI={10.1103/physrevlett.106.181301},
   number={18},
   journal={Physical Review Letters},
   publisher={American Physical Society (APS)},
   author={Barnaby, Neil and Peloso, Marco},
   year={2011},
   month=May }

@article{Durrer_2024,
   title={Scalar perturbations from inflation in the presence of gauge fields},
   volume={110},
   ISSN={2470-0029},
   url={http://dx.doi.org/10.1103/PhysRevD.110.043533},
   DOI={10.1103/physrevd.110.043533},
   number={4},
   journal={Physical Review D},
   publisher={American Physical Society (APS)},
   author={Durrer, R. and von Eckardstein, R. and Garg, Deepen and Schmitz, K. and Sobol, O. and Vilchinskii, S.},
   year={2024},
   month=Aug }

@article{Gaspard_2018,
   title={Connection formulas between Coulomb wave functions},
   volume={59},
   ISSN={1089-7658},
   url={http://dx.doi.org/10.1063/1.5054368},
   DOI={10.1063/1.5054368},
   number={11},
   journal={Journal of Mathematical Physics},
   publisher={AIP Publishing},
   author={Gaspard, David},
   year={2018},
   month=Nov }

@article{STAROBINSKY,
title = {A new type of isotropic cosmological models without singularity},
journal = {Physics Letters B},
volume = {91},
number = {1},
pages = {99-102},
year = {1980},
issn = {0370-2693},
doi = {https://doi.org/10.1016/0370-2693(80)90670-X},
url = {https://www.sciencedirect.com/science/article/pii/037026938090670X},
author = {A.A. Starobinsky}
}

@article{Guth,
  title = {Inflationary universe: A possible solution to the horizon and flatness problems},
  author = {Guth, Alan H.},
  journal = {Phys. Rev. D},
  volume = {23},
  issue = {2},
  pages = {347--356},
  numpages = {0},
  year = {1981},
  month = {Jan},
  publisher = {American Physical Society},
  doi = {10.1103/PhysRevD.23.347},
  url = {https://link.aps.org/doi/10.1103/PhysRevD.23.347}
}

@article{Mukhanov,
    author = "Mukhanov, Viatcheslav F. and Chibisov, G. V.",
    title = "{Quantum Fluctuations and a Nonsingular Universe}",
    journal = "JETP Lett.",
    volume = "33",
    pages = "532--535",
    year = "1981"
}

@article{Cahn_23,
  title = {Test for Cosmological Parity Violation Using the 3D Distribution of Galaxies},
  author = {Cahn, Robert N. and Slepian, Zachary and Hou, Jiamin},
  journal = {Phys. Rev. Lett.},
  volume = {130},
  issue = {20},
  pages = {201002},
  numpages = {6},
  year = {2023},
  month = {May},
  publisher = {American Physical Society},
  doi = {10.1103/PhysRevLett.130.201002},
  url = {https://link.aps.org/doi/10.1103/PhysRevLett.130.201002}
}

@article{Hou_2023,
   title={Measurement of parity-odd modes in the large-scale 4-point correlation function of Sloan Digital Sky Survey Baryon Oscillation Spectroscopic Survey twelfth data release CMASS and LOWZ galaxies},
   volume={522},
   ISSN={1365-2966},
   url={http://dx.doi.org/10.1093/mnras/stad1062},
   DOI={10.1093/mnras/stad1062},
   number={4},
   journal={Monthly Notices of the Royal Astronomical Society},
   publisher={Oxford University Press (OUP)},
   author={Hou, Jiamin and Slepian, Zachary and Cahn, Robert N},
   year={2023},
   month=May, pages={5701–5739} }

@article{Freese,
  title = {Natural inflation with pseudo Nambu-Goldstone bosons},
  author = {Freese, Katherine and Frieman, Joshua A. and Olinto, Angela V.},
  journal = {Phys. Rev. Lett.},
  volume = {65},
  issue = {26},
  pages = {3233--3236},
  numpages = {0},
  year = {1990},
  month = {Dec},
  publisher = {American Physical Society},
  doi = {10.1103/PhysRevLett.65.3233},
  url = {https://link.aps.org/doi/10.1103/PhysRevLett.65.3233}
}

@article{Barnaby_2011_2,
   title={Large Non-Gaussianity in Axion Inflation},
   volume={106},
   ISSN={1079-7114},
   url={http://dx.doi.org/10.1103/PhysRevLett.106.181301},
   DOI={10.1103/physrevlett.106.181301},
   number={18},
   journal={Physical Review Letters},
   publisher={American Physical Society (APS)},
   author={Barnaby, Neil and Peloso, Marco},
   year={2011},
   month=May }

@article{Frieman_2008,
   title={Dark Energy and the Accelerating Universe},
   volume={46},
   ISSN={1545-4282},
   url={http://dx.doi.org/10.1146/annurev.astro.46.060407.145243},
   DOI={10.1146/annurev.astro.46.060407.145243},
   number={1},
   journal={Annual Review of Astronomy and Astrophysics},
   publisher={Annual Reviews},
   author={Frieman, Joshua A. and Turner, Michael S. and Huterer, Dragan},
   year={2008},
   month=Sept, pages={385–432} }

@book{Baumann,
    author = "Baumann, Daniel",
    title = "{Cosmology}",
    doi = "10.1017/9781108937092",
    isbn = "978-1-108-93709-2, 978-1-108-83807-8",
    publisher = "Cambridge University Press",
    month = "7",
    year = "2022"
}

@ARTICLE{Bond_87,
       author = {{Bond}, J.~R. and {Efstathiou}, G.},
        title = "{The statistics of cosmic background radiation fluctuations}",
      journal = {Monthly Notices of the Royal Astronomical Society},
         year = 1987,
        month = jun,
       volume = {226},
        pages = {655-687},
          doi = {10.1093/mnras/226.3.655},
       adsurl = {https://ui.adsabs.harvard.edu/abs/1987MNRAS.226..655B}
}

@article{Hou_2026,
   title={Parity-odd four-point correlation function from the DESI data release 1 luminous red galaxy sample},
   volume={113},
   ISSN={2470-0029},
   url={http://dx.doi.org/10.1103/2dmd-hyt1},
   DOI={10.1103/2dmd-hyt1},
   number={10},
   journal={Physical Review D},
   publisher={American Physical Society (APS)},
   author={Hou, J. and Cahn, R. N. and Aguilar, J. and Ahlen, S. and Bianchi, D. and Brooks, D. and Claybaugh, T. and Doel, P. and Forero-Romero, J. E. and Gaztañaga, E. and Le Guillou, L. and Gutierrez, G. and Howlett, C. and Huterer, D. and Ishak, M. and Joyce, R. and Kremin, A. and Lahav, O. and Lamman, C. and Landriau, M. and de la Macorra, A. and Miquel, R. and Nadathur, S. and Niz, G. and Percival, W. J. and Prada, F. and Pérez-Ràfols, I. and Rossi, G. and Sanchez, E. and Schlegel, D. and Schubnell, M. and Seo, H. and Silber, J. and Sprayberry, D. and Tarlé, G. and Weaver, B. A. and Zou, H.},
   year={2026},
   month=May }

@misc{slepian2025,
      title={Measurement of Parity-Violating Modes of the Dark Energy Spectroscopic Instrument (DESI) Year 1 Luminous Red Galaxies' 4-Point Correlation Function}, 
      author={Zachary Slepian and Alex Krolewski and Alessandro Greco and Simon May and William Ortola Leonard and Farshad Kamalinejad and Jessica Chellino and Matthew Reinhard and Elena Fernandez and Francisco Prada and Steven Ahlen and Davide Bianchi and David Brooks and Todd Claybaugh and Axel de la Macorra and Arnaud de Mattia and Biprateep Dey and Peter Doel and Enrique Gaztanaga and Gaston Gutierrez and Klaus Honscheid and Dragan Huterer and Dick Joyce and Robert Kehoe and David Kirkby and Theodore Kisner and Martin Landriau and Laurent Le Guillou and Marc Manera and Aaron Meisner and Ramon Miquel and Seshadri Nadathur and Will Percival and Ashley Ross and Eusebio Sanchez and David Schlegel and Michael Schubnell and Hee-Jong Seo and Joseph Silber and David Sprayberry and Gregory Tarle},
      year={2025},
      eprint={2508.09133},
      archivePrefix={arXiv},
      primaryClass={astro-ph.CO},
      url={https://arxiv.org/abs/2508.09133}, 
}

@article{Niu_2023,
   title={Parity-odd and even trispectrum from axion inflation},
   volume={2023},
   ISSN={1475-7516},
   url={http://dx.doi.org/10.1088/1475-7516/2023/05/018},
   DOI={10.1088/1475-7516/2023/05/018},
   number={05},
   journal={Journal of Cosmology and Astroparticle Physics},
   publisher={IOP Publishing},
   author={Niu, Xuce and Rahat, Moinul Hossain and Srinivasan, Karthik and Xue, Wei},
   year={2023},
   month=May, pages={018} }

@article{Caravano_2023,
   title={Lattice simulations of axion-U(1) inflation},
   volume={108},
   ISSN={2470-0029},
   url={http://dx.doi.org/10.1103/PhysRevD.108.043504},
   DOI={10.1103/physrevd.108.043504},
   number={4},
   journal={Physical Review D},
   publisher={American Physical Society (APS)},
   author={Caravano, Angelo and Komatsu, Eiichiro and Lozanov, Kaloian D. and Weller, Jochen},
   year={2023},
   month=Aug }

@article{Figueroa_2025,
   title={Nonlinear dynamics of axion inflation: A detailed lattice study},
   volume={111},
   ISSN={2470-0029},
   url={http://dx.doi.org/10.1103/PhysRevD.111.063545},
   DOI={10.1103/physrevd.111.063545},
   number={6},
   journal={Physical Review D},
   publisher={American Physical Society (APS)},
   author={Figueroa, Daniel G. and Lizarraga, Joanes and Loayza, Nicolás and Urio, Ander and Urrestilla, Jon},
   year={2025},
   month=Mar }

@article{Jamieson_2025,
   title={Primordial power spectrum and bispectrum from lattice simulations of axion-U(1) inflation},
   volume={112},
   ISSN={2470-0029},
   url={http://dx.doi.org/10.1103/3zml-71jd},
   DOI={10.1103/3zml-71jd},
   number={10},
   journal={Physical Review D},
   publisher={American Physical Society (APS)},
   author={Jamieson, Drew and Caravano, Angelo and Komatsu, Eiichiro},
   year={2025},
   month=Nov }

@article{Ashtekar_2020,
   title={Emergence of classical behavior in the early Universe},
   volume={102},
   ISSN={2470-0029},
   url={http://dx.doi.org/10.1103/PhysRevD.102.023512},
   DOI={10.1103/physrevd.102.023512},
   number={2},
   journal={Physical Review D},
   publisher={American Physical Society (APS)},
   author={Ashtekar, Abhay and Corichi, Alejandro and Kesavan, Aruna},
   year={2020},
   month=July }

@article{Fujita_2024,
   title={Parity-violating scalar trispectrum from a rolling axion during inflation},
   volume={2024},
   ISSN={1475-7516},
   url={http://dx.doi.org/10.1088/1475-7516/2024/05/127},
   DOI={10.1088/1475-7516/2024/05/127},
   number={05},
   journal={Journal of Cosmology and Astroparticle Physics},
   publisher={IOP Publishing},
   author={Fujita, Tomohiro and Murata, Tomoaki and Obata, Ippei and Shiraishi, Maresuke},
   year={2024},
   month=May, pages={127} }

@article{Taruya_2018,
   title={Grid-based calculation for perturbation theory of large-scale structure},
   volume={98},
   ISSN={2470-0029},
   url={http://dx.doi.org/10.1103/PhysRevD.98.103532},
   DOI={10.1103/physrevd.98.103532},
   number={10},
   journal={Physical Review D},
   publisher={American Physical Society (APS)},
   author={Taruya, Atsushi and Nishimichi, Takahiro and Jeong, Donghui},
   year={2018},
   month=Nov }

@article{Springel_2021,
  author    = {Springel, Volker and Pakmor, R{\"u}diger and Zier, Oliver and Reinecke, Martin},
  title     = {Simulating cosmic structure formation with the {GADGET}-4 code},
  journal   = {Monthly Notices of the Royal Astronomical Society},
  volume    = {506},
  number    = {2},
  pages     = {2871--2949},
  year      = {2021},
  month     = jul,
  doi       = {10.1093/mnras/stab1855},
  url       = {https://doi.org/10.1093/mnras/stab1855}
}

@article{Krolewski_2024,
   title={No evidence for parity violation in BOSS},
   volume={2024},
   ISSN={1475-7516},
   url={http://dx.doi.org/10.1088/1475-7516/2024/08/044},
   DOI={10.1088/1475-7516/2024/08/044},
   number={08},
   journal={Journal of Cosmology and Astroparticle Physics},
   publisher={IOP Publishing},
   author={Krolewski, Alex and May, Simon and Smith, Kendrick and Hopkins, Hans},
   year={2024},
   month=Aug, pages={044} }

@article {dealising,
      author = "Steven A.  Orszag",
      title = "On the Elimination of Aliasing in Finite-Difference Schemes by Filtering High-Wavenumber Components",
      journal = "Journal of Atmospheric Sciences",
      year = "1971",
      publisher = "American Meteorological Society",
      address = "Boston MA, USA",
      volume = "28",
      number = "6",
      doi = "10.1175/1520-0469(1971)028<1074:OTEOAI>2.0.CO;2",
      pages=      "1074 - 1074",
      url = "https://journals.ametsoc.org/view/journals/atsc/28/6/1520-0469_1971_028_1074_oteoai_2_0_co_2.xml"
}

@article{Taruya_2022,
   title={Grid-based calculations of redshift-space matter fluctuations from perturbation theory: UV sensitivity and convergence at the field level},
   volume={105},
   ISSN={2470-0029},
   url={http://dx.doi.org/10.1103/PhysRevD.105.103507},
   DOI={10.1103/physrevd.105.103507},
   number={10},
   journal={Physical Review D},
   publisher={American Physical Society (APS)},
   author={Taruya, Atsushi and Nishimichi, Takahiro and Jeong, Donghui},
   year={2022},
   month=May }

@article{Taruya_2021,
   title={Covariance of the matter power spectrum including the survey window function effect: 
<mml:math xmlns:mml=“http://www.w3.org/1998/Math/MathML” display=“inline”><mml:mi>N</mml:mi></mml:math>
-body simulations versus fifth-order perturbation theory on grids},
   volume={103},
   ISSN={2470-0029},
   url={http://dx.doi.org/10.1103/PhysRevD.103.023501},
   DOI={10.1103/physrevd.103.023501},
   number={2},
   journal={Physical Review D},
   publisher={American Physical Society (APS)},
   author={Taruya, Atsushi and Nishimichi, Takahiro and Jeong, Donghui},
   year={2021},
   month=Jan }

@article{Planck_2018,
   title={<i>Planck</i>
                    2018 results: X. Constraints on inflation},
   volume={641},
   ISSN={1432-0746},
   url={http://dx.doi.org/10.1051/0004-6361/201833887},
   DOI={10.1051/0004-6361/201833887},
   journal={Astronomy \& Astrophysics},
   publisher={EDP Sciences},
   author={ and Akrami, Y. and Arroja, F. and Ashdown, M. and Aumont, J. and Baccigalupi, C. and Ballardini, M. and Banday, A. J. and Barreiro, R. B. and Bartolo, N. and Basak, S. and Benabed, K. and Bernard, J.-P. and Bersanelli, M. and Bielewicz, P. and Bock, J. J. and Bond, J. R. and Borrill, J. and Bouchet, F. R. and Boulanger, F. and Bucher, M. and Burigana, C. and Butler, R. C. and Calabrese, E. and Cardoso, J.-F. and Carron, J. and Challinor, A. and Chiang, H. C. and Colombo, L. P. L. and Combet, C. and Contreras, D. and Crill, B. P. and Cuttaia, F. and de Bernardis, P. and de Zotti, G. and Delabrouille, J. and Delouis, J.-M. and Di Valentino, E. and Diego, J. M. and Donzelli, S. and Doré, O. and Douspis, M. and Ducout, A. and Dupac, X. and Dusini, S. and Efstathiou, G. and Elsner, F. and Enßlin, T. A. and Eriksen, H. K. and Fantaye, Y. and Fergusson, J. and Fernandez-Cobos, R. and Finelli, F. and Forastieri, F. and Frailis, M. and Franceschi, E. and Frolov, A. and Galeotta, S. and Galli, S. and Ganga, K. and Gauthier, C. and Génova-Santos, R. T. and Gerbino, M. and Ghosh, T. and González-Nuevo, J. and Górski, K. M. and Gratton, S. and Gruppuso, A. and Gudmundsson, J. E. and Hamann, J. and Handley, W. and Hansen, F. K. and Herranz, D. and Hivon, E. and Hooper, D. C. and Huang, Z. and Jaffe, A. H. and Jones, W. C. and Keihänen, E. and Keskitalo, R. and Kiiveri, K. and Kim, J. and Kisner, T. S. and Krachmalnicoff, N. and Kunz, M. and Kurki-Suonio, H. and Lagache, G. and Lamarre, J.-M. and Lasenby, A. and Lattanzi, M. and Lawrence, C. R. and Le Jeune, M. and Lesgourgues, J. and Levrier, F. and Lewis, A. and Liguori, M. and Lilje, P. B. and Lindholm, V. and López-Caniego, M. and Lubin, P. M. and Ma, Y.-Z. and Macías-Pérez, J. F. and Maggio, G. and Maino, D. and Mandolesi, N. and Mangilli, A. and Marcos-Caballero, A. and Maris, M. and Martin, P. G. and Martínez-González, E. and Matarrese, S. and Mauri, N. and McEwen, J. D. and Meerburg, P. D. and Meinhold, P. R. and Melchiorri, A. and Mennella, A. and Migliaccio, M. and Mitra, S. and Miville-Deschênes, M.-A. and Molinari, D. and Moneti, A. and Montier, L. and Morgante, G. and Moss, A. and Münchmeyer, M. and Natoli, P. and Nørgaard-Nielsen, H. U. and Pagano, L. and Paoletti, D. and Partridge, B. and Patanchon, G. and Peiris, H. V. and Perrotta, F. and Pettorino, V. and Piacentini, F. and Polastri, L. and Polenta, G. and Puget, J.-L. and Rachen, J. P. and Reinecke, M. and Remazeilles, M. and Renzi, A. and Rocha, G. and Rosset, C. and Roudier, G. and Rubiño-Martín, J. A. and Ruiz-Granados, B. and Salvati, L. and Sandri, M. and Savelainen, M. and Scott, D. and Shellard, E. P. S. and Shiraishi, M. and Sirignano, C. and Sirri, G. and Spencer, L. D. and Sunyaev, R. and Suur-Uski, A.-S. and Tauber, J. A. and Tavagnacco, D. and Tenti, M. and Toffolatti, L. and Tomasi, M. and Trombetti, T. and Valiviita, J. and Van Tent, B. and Vielva, P. and Villa, F. and Vittorio, N. and Wandelt, B. D. and Wehus, I. K. and White, S. D. M. and Zacchei, A. and Zibin, J. P. and Zonca, A.},
   year={2020},
   month=Sept, pages={A10} }

@ARTICLE{Corba_2026,
       author = {{Corb{\`a}}, Sofia P.},
        title = "{Gravitational wave anisotropies from axion inflation}",
      journal = {Journal of Cosmology and Astroparticle Physics},
         year = 2026,
        month = jan,
       volume = {2026},
       number = {1},
          eid = {029},
        pages = {029},
          doi = {10.1088/1475-7516/2026/01/029},
archivePrefix = {arXiv},
       eprint = {2504.13156},
 primaryClass = {astro-ph.CO},
       adsurl = {https://ui.adsabs.harvard.edu/abs/2026JCAP...01..029C}
}

@article{Sorbo_2011,
   title={Parity violation in the Cosmic Microwave Background from a pseudoscalar inflaton},
   volume={2011},
   ISSN={1475-7516},
   url={http://dx.doi.org/10.1088/1475-7516/2011/06/003},
   DOI={10.1088/1475-7516/2011/06/003},
   number={06},
   journal={Journal of Cosmology and Astroparticle Physics},
   publisher={IOP Publishing},
   author={Sorbo, Lorenzo},
   year={2011},
   month=June, pages={003–003} }

@article{Barnaby_2012,
   title={Gauge field production in axion inflation: Consequences for monodromy, non-Gaussianity in the CMB, and gravitational waves at interferometers},
   volume={85},
   ISSN={1550-2368},
   url={http://dx.doi.org/10.1103/PhysRevD.85.023525},
   DOI={10.1103/physrevd.85.023525},
   number={2},
   journal={Physical Review D},
   publisher={American Physical Society (APS)},
   author={Barnaby, Neil and Pajer, Enrico and Peloso, Marco},
   year={2012},
   month=Jan }

@article{Cook_2012,
   title={Particle production during inflation and gravitational waves detectable by ground-based interferometers},
   volume={85},
   ISSN={1550-2368},
   url={http://dx.doi.org/10.1103/PhysRevD.85.023534},
   DOI={10.1103/physrevd.85.023534},
   number={2},
   journal={Physical Review D},
   publisher={American Physical Society (APS)},
   author={Cook, Jessica L. and Sorbo, Lorenzo},
   year={2012},
   month=Jan }

@article{Anber_2012,
   title={Non-Gaussianities and chiral gravitational waves in natural steep inflation},
   volume={85},
   ISSN={1550-2368},
   url={http://dx.doi.org/10.1103/PhysRevD.85.123537},
   DOI={10.1103/physrevd.85.123537},
   number={12},
   journal={Physical Review D},
   publisher={American Physical Society (APS)},
   author={Anber, Mohamed M. and Sorbo, Lorenzo},
   year={2012},
   month=June }

@article{Linde_2013,
   title={Gauge field production in supergravity inflation: Local non-Gaussianity and primordial black holes},
   volume={87},
   ISSN={1550-2368},
   url={http://dx.doi.org/10.1103/PhysRevD.87.103506},
   DOI={10.1103/physrevd.87.103506},
   number={10},
   journal={Physical Review D},
   publisher={American Physical Society (APS)},
   author={Linde, Andrei and Mooij, Sander and Pajer, Enrico},
   year={2013},
   month=May }

@article{Meerburg_2013,
   title={Observational constraints on gauge field production in axion inflation},
   volume={2013},
   ISSN={1475-7516},
   url={http://dx.doi.org/10.1088/1475-7516/2013/02/017},
   DOI={10.1088/1475-7516/2013/02/017},
   number={02},
   journal={Journal of Cosmology and Astroparticle Physics},
   publisher={IOP Publishing},
   author={Meerburg, P.D and Pajer, E},
   year={2013},
   month=Feb, pages={017–017} }

@article{Ferreira_2014,
author = {Zambujal Ferreira, Ricardo and Sloth, Martin},
year = {2014},
month = {12},
pages = {},
title = {Universal constraints on axions from inflation},
volume = {2014},
journal = {Journal of High Energy Physics},
doi = {10.1007/JHEP12(2014)139}
}

@article{Garc_a_Bellido_2016,
   title={Gravitational waves at interferometer scales and primordial black holes in axion inflation},
   volume={2016},
   ISSN={1475-7516},
   url={http://dx.doi.org/10.1088/1475-7516/2016/12/031},
   DOI={10.1088/1475-7516/2016/12/031},
   number={12},
   journal={Journal of Cosmology and Astroparticle Physics},
   publisher={IOP Publishing},
   author={García-Bellido, Juan and Peloso, Marco and Unal, Caner},
   year={2016},
   month=Dec, pages={031–031} }

@article{Domcke_2017,
   title={PBH dark matter from axion inflation},
   volume={2017},
   ISSN={1475-7516},
   url={http://dx.doi.org/10.1088/1475-7516/2017/07/048},
   DOI={10.1088/1475-7516/2017/07/048},
   number={07},
   journal={Journal of Cosmology and Astroparticle Physics},
   publisher={IOP Publishing},
   author={Domcke, Valerie and Muia, Francesco and Pieroni, Mauro and Witkowski, Lukas T.},
   year={2017},
   month=July, pages={048–048} }

@article{_zsoy_2023,
   title={Inflation and Primordial Black Holes},
   volume={9},
   ISSN={2218-1997},
   url={http://dx.doi.org/10.3390/universe9050203},
   DOI={10.3390/universe9050203},
   number={5},
   journal={Universe},
   publisher={MDPI AG},
   author={Özsoy, Ogan and Tasinato, Gianmassimo},
   year={2023},
   month=Apr, pages={203} }

@misc{alam2024,
      title={CMB Constraints on Natural Inflation with Gauge Field Production}, 
      author={Khursid Alam and Koushik Dutta and Nur Jaman},
      year={2024},
      eprint={2405.10155},
      archivePrefix={arXiv},
      primaryClass={astro-ph.CO},
      url={https://arxiv.org/abs/2405.10155}, 
}

@article{Niu_2023_2,
   title={Gravitational wave probes of massive gauge bosons at the cosmological collider},
   volume={2023},
   ISSN={1475-7516},
   url={http://dx.doi.org/10.1088/1475-7516/2023/02/013},
   DOI={10.1088/1475-7516/2023/02/013},
   number={02},
   journal={Journal of Cosmology and Astroparticle Physics},
   publisher={IOP Publishing},
   author={Niu, Xuce and Rahat, Moinul Hossain and Srinivasan, Karthik and Xue, Wei},
   year={2023},
   month=Feb, pages={013} }
\end{document}